\documentclass[aps,prd,fleqn,floatfix,nofootinbib,showpacs,preprintnumbers]{revtex4}
\usepackage{amsmath,amssymb,amsfonts}
\usepackage{graphicx}% Include figure files
\usepackage{bm}% bold math

\begin {document}

\title{The Transverse-momentum-dependent Parton Distribution Function
and Jet Transport in Medium}

\author{Zuo-tang Liang$^a$, Xin-Nian Wang$^{b}$ and Jian Zhou$^{a,b}$}
\address{$^a$Department of Physics, Shandong University, Jinan,
Shandong 250100, China}

\address{$^b$Nuclear Science Division, MS 70R0319, Lawrence Berkeley National
                   Laboratory, Berkeley, CA 94720}

\date{December 8, 2007}

\preprint {LBNL-63708}

\begin{abstract}
We show that the gauge-invariant transverse-momentum-dependent (TMD)
quark distribution function can be expressed as
a sum of all higher-twist collinear parton matrix elements
in terms of a transport operator. From such a general expression, we
derive the nuclear broadening of the transverse momentum
distribution.  Under the maximal two-gluon correlation approximation,
in which all higher-twist nuclear multiple-parton correlations with
the leading nuclear enhancement are given by products of twist-two
nucleon parton distributions, we find the nuclear transverse momentum
distribution as a convolution of a Gaussian distribution and
the nucleon TMD quark distribution. The width of the Gaussian,
or the mean total transverse momentum broadening squared, is given
by the path integral of the quark transport parameter $\hat q_F$ which
can also be expressed in a gauge invariant form and is given by the
gluon distribution density in the nuclear medium. We further show that
contributions from higher-twist nucleon gluon distributions can be resummed
under the extended adjoint two-gluon correlation approximation and
the nuclear transverse momentum distribution can be expressed in terms
of a transverse scale dependent quark transport parameter or gluon distribution
density. We extend the study to hot medium and compare to dipole
model approximation and ${\cal N}=4$ Supersymmetric 
Yang-Mills (SYM) theory in the strong coupling limit.
We find that multiple gluon correlations become important in
the strongly coupled system such as ${\cal N}=4$ SYM plasma.

\end{abstract}

\pacs{13.85.Hd, 25.75.Bh,11.80.La}

\maketitle

\section {Introduction}

The success of perturbative QCD (pQCD) in describing hard
processes in hadronic interactions relies on the factorization
theorem \cite{Collins:1981uw} that separates the coherent long
distance interaction between projectile and target from
the incoherent short distance interactions. The physical
observables such as cross sections of deeply inelastic scattering
(DIS) and Drell-Yan (DY) dilepton production can be expressed as a
convolution of hard partonic scattering cross sections, parton
distribution functions and parton fragmentation functions. 
The hard partonic parts are calculable in
a perturbative expansion in the strong coupling constant
$\alpha_s(Q^2)$ which becomes small large
momentum scale $Q^2$ of the hard processes
\cite{Gross:1973id,Politzer:1973fx}. Though the parton
distribution and fragmentation functions are not calculable in
pQCD since they involve long distance interaction, they are
universal and independent of the specific partonic hard processes.
Therefore, they can be measured in one hard process and then
applied to another, therein lies the predictable power of pQCD.

The most practiced factorization scheme is collinear factorization in which
one integrates out the transverse momentum of the initial (final) parton
up to a factorization scale and the final observables will only depend on
the transverse-momentum-integrated or collinear factorized parton
distribution (fragmentation) functions. Such a proof of factorization
has also been extended to semi-inclusive processes \cite{Bauer:2002nz,Ji:2004wu}
that involve finite transverse momentum of the final hadron or dilepton
with the introduction of transverse-momentum-dependent (TMD) parton distribution
and fragmentation functions. The final observables can be expressed
as a convolution of collinear hard parts (setting the initial parton
transverse momenta to zero) and TMD parton distribution and fragmentation
functions \cite{Liang:2006wp}. Such TMD parton distribution and
fragmentation functions are important for the study of hadronic interactions
with singly or doubly polarized beams, such as single-spin asymmetry in
semi-inclusive processes in DIS
(SIDIS) \cite{Sivers:1989cc,Brodsky:2002cx,Collins:2002kn}
and proton-proton scattering.

\begin{figure}
\centerline{\includegraphics[width=9cm]{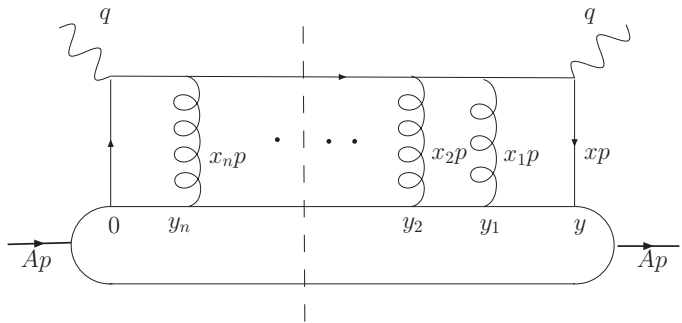}}
\caption{Multiple soft gluon interaction between the struck quark and the
remnant of the target nucleus in DIS.}
\label{fig1}
\end{figure}

In the proof of factorization \cite{Belitsky:2002sm}, in DIS off a
nucleon or nucleus target for example, one important step is to
eikonalize all soft interactions between the struck quark and the
target remnant as shown in Fig.~\ref{fig1}. The summation of these
soft gluon interactions gives rise to a definition of TMD quark
distribution function in a nucleus,
\begin{eqnarray}
f_q^A(x,\vec k_\perp) = \int \frac{dy^-}{2\pi} \frac{d^2y_\perp}{(2\pi)^2}
e^{ixp^+y^- -i\vec k_\perp\cdot \vec y_\perp}
\langle A \mid \bar\psi(0,\vec 0_\perp)\frac{\gamma^+}{2}
{\cal L}_{\rm TMD}(0,y) \psi(y^-,\vec y_\perp)
\mid A \rangle ,
\label{tmd0}
\end{eqnarray}
where,
\begin{equation}
{\cal L}_{\rm TMD}(0,y)\equiv
{\cal L}^\dagger_\parallel(-\infty,0;\vec 0_\perp)
{\cal L}^\dagger_\perp(-\infty;\vec y_\perp,\vec 0_\perp)
{\cal L}_\parallel(-\infty,y^-;\vec y_\perp)
\end{equation}
is the complete gauge link in TMD quark distribution function that
contains both the transverse \cite{Belitsky:2002sm}
\begin{equation}
{\cal L}_\perp(-\infty;\vec y_\perp,\vec 0_\perp)
\equiv P\exp\left[-ig\int_{\vec 0_\perp}^{\vec y_\perp} d\vec\xi_\perp\cdot
\vec A_\perp(-\infty,\vec\xi_\perp)\right]
\end{equation}
and longitudinal gauge link \cite{Liang:2006wp}
\begin{equation}
{\cal L}_\parallel (-\infty,y^-;\vec y_\perp)\equiv
P\exp\left[-ig\int^{-\infty}_{y^-} d\xi^- A_+(\xi^-,\vec y_\perp)\right].
\end{equation}
The above gauge links for quark propagation are defined in the
fundamental representation, $A_\mu=A^a_\mu T^a (a=1-8)$. Note that
in our convention, the quark is produced at $(y^-, \vec y_\perp)$
[ or $(0, \vec 0_\perp)$] and propagates toward $(-\infty,\vec
y_\perp)$ [$(-\infty,\vec 0_\perp)$]. The path ordering along the
light-cone is then defined from $y^-$ to $-\infty$.

Even though the parton distribution in Eq.~(\ref{tmd0}) describes the
probability to find a quark with momentum fraction $x$ and transverse
momentum $\vec k_\perp$ in a nucleus, it also contains information about
final state interaction with the target remnant encoded through the
gauge links. These gauge links are not only crucial to ensure the gauge
invariance of the TMD parton distribution functions in both light-cone and
covariant gauge but also lead to physical consequences such as single-spin
asymmetry in semi-inclusive DIS and Drell-Yan
process \cite{Brodsky:2002cx,Belitsky:2002sm,Ji:2006ub}. For DIS
off a nucleus target, they should also contain information about
transverse momentum broadening of the struck quark due to multiple
scattering inside the nucleus. The main purpose of this work is to
study nuclear transverse momentum broadening in DIS from the TMD
quark distribution functions and extend the result to the case
of quark or jet propagation in a thermal medium as in high-energy
heavy-ion collisions.

In the study of nuclear matter and quark-gluon plasma in high
energy lepton-nucleus, hadron-nucleus and nucleus-nucleus
collisions, jet transverse momentum broadening plays a crucial
role in unravelling the medium properties through modification of
the final jet or hadron spectra (jet quenching) due to final state
interaction between the energetic partons and the nuclear or hot
medium \cite{Wang:1991xy,Wang:2002ri}. Current phenomenological
studies of experimental data
\cite{Vitev:2002pf,Wang:2003mm,Eskola:2004cr,Turbide:2005fk,Majumder:2007ae}
on jet quenching rely on pQCD calculations of the parton energy loss
or modification of the parton fragmentation functions due to gluon
radiation induced by multiple scattering during the parton
propagation in the medium. One important parameter that controls
parton energy loss or medium modification of the jet fragmentation
function is the jet transport parameter $\hat q$ or transverse
momentum broadening squared per unit of propagation length
\cite{Gyulassy:1993hr,Zakharov:1996fv,
Baier:1996sk,Gyulassy:2000er,Wiedemann:2000za,Guo:2000nz,Wang:2001ifa}.
Therefore, calculation and measurement of the jet transport
parameter is an important step toward understanding the intrinsic
properties of the QCD medium.

There are many calculations of the jet transverse momentum
broadening. Early studies dealt with the mean average transport 
parameter \cite{Baier:1996sk,Guo:1998rd} under one particular 
gauge without apparent
guaranty of gauge invariance. Transverse momentum broadening in
Drell-Yan pair production in proton-nucleus scattering \cite{Fries:2002mu}
and a recent calculation of nuclear transverse momentum broadening
distribution in DIS \cite{Majumder:2007hx} were
obtained  by a direct summation of multiple
scattering in covariant gauge, again without apparent gauge
invariance in the final result. One approach to the parton
propagation in medium with a gauge (mostly) invariant framework is
the Wilson line formulation of multiple parton scattering
\cite{Wiedemann:2000za} that resembles the longitudinal gauge link
in Eq.~(\ref{tmd0}), but again in the covariant gauge. Since
resummation of all soft gluon interactions is crucial for the
gauge invariant form of the TMD parton distribution function in
Eq.(\ref{tmd0}), it must contain the nuclear transverse momentum
broadening due to multiple parton scattering in nuclei. One should
be able to derive a completely gauge invariant form of the nuclear transverse
momentum broadening from the nuclear TMD quark distribution
function. This is what we will prove in this paper. We will derive the nuclear
broadening of the transverse momentum distribution simply from the gauge invariant
form of TMD quark distribution function in Eq.(\ref{tmd0}) with a given nuclear
distribution function. The broadened distribution will have a Gaussian form as
found in earlier studies, considering only two-gluon field correlations in the
nucleons. However, the broadening parameter in our derivation will have 
an explicit gauge invariant form. We will also show that contributions from 
higher-twist multi-gluon correlations in a nucleon can be resummed to give 
a nuclear TMD quark distribution that depends on the transverse scale 
dependent  gluon distribution density inside the nucleus.

We will briefly summarize our main results here.
To calculate the nuclear transverse momentum
broadening including all higher-twist nuclear parton matrix elements,
we first express the gauge-invariant
TMD quark distribution function in a nucleus as a sum of
collinear higher-twist nuclear parton matrix elements which can
be exponentiated to give,
\begin{eqnarray}
f_q^A(x_B,\vec k_\perp)&=&\int \frac{dy^-}{2\pi}e^{ixp^+y^-}
\langle A \mid \bar\psi(0,\vec 0_\perp)\frac{\gamma^+}{2}
{\cal L}_\parallel(0,y^-;\vec 0_\perp)
e^{\vec W_\perp(y^-\! ,\, \vec 0_\perp) \cdot\vec\nabla_{k_\perp}}
\psi(y^-,\vec 0_\perp)\mid A \rangle
\delta^{(2)}(\vec k_\perp),
\end{eqnarray}
where the transport operator $\vec W_\perp(y^-,\vec y_\perp)$ is defined as
\begin{eqnarray}
\vec W_\perp(y^-\! ,\,\vec y_\perp)&\equiv&
i\vec D_\perp(y^-,\vec y_\perp)
+g\int_{-\infty}^{y^-}d\xi^-{\cal L}^\dagger_\parallel(\xi,y^-,\vec y_\perp)
\vec F_{+\perp}(\xi^-,\vec y_\perp)
{\cal L}_\parallel(\xi^-,y^-;\vec y_\perp),
\end{eqnarray}
and $\vec D_\perp(y^-,\vec y_\perp)=\vec\partial_\perp
+ig\vec A_\perp(y^-,\vec y_\perp)$
is the covariant derivative. The transport operator is
supposed to operate on both the quark field and the gluon fields
within itself and the transverse coordinate is set 
to $\vec y_\perp=\vec 0_\perp$ after the operation.

With a maximal two-gluon correlation approximation,
the high-twist multi-parton correlations in a large and weakly
bound nucleus can be expressed as products, which have the maximum 
nuclear size (or medium length) dependence, of twist-two nucleon
gluon distribution functions. The leading contribution to
the nuclear TMD quark distribution is shown to have a simple form,
\begin{equation}
f_q^A(x_B,\vec k_\perp)=A \exp\left[\int d\xi^-_N \hat q_F(\xi_N)
\frac{\nabla_{k_\perp}^2}{4}\right] f_q^N(x_B,\vec k_\perp)
=\frac{A}{\pi \Delta_{2F}}
\int d^2\ell_\perp e^{-(\vec k_\perp-\vec\ell_\perp)^2/
\Delta_{2F}}f_q^N(x,\vec \ell_\perp),
\end{equation}
with $\Delta_{2F}=\int d\xi_N^-\hat q_F(\xi_N)$ as the total
transverse momentum broadening squared. Such a distribution also
satisfies a 2-D diffusion equation in transverse momentum $\vec
k_\perp$ with the diffusion constant given by $\hat q_F$, the
so-called (twist-2) quark transport parameter,
\begin{eqnarray}
\hat q_F(\xi_N)
&=&\frac{2\pi^2\alpha_s}{N_c}\rho^A_N(\xi_N)
\int \frac{d\xi^-}{2\pi p^+}(-2)
\langle N \mid
{\rm Tr} \left[
F_{+\sigma}(0,\vec 0_\perp){\cal L}_\parallel(0,\xi^-;\vec 0_\perp)
F_+^{\,\,\sigma}(\xi^-,\vec 0_\perp)
{\cal L}_\parallel^\dagger(0,\xi^-;\vec 0_\perp)\right]
\mid N \rangle,
\end{eqnarray}
which is related to the twist-two collinear nucleon gluon distribution
function. It is also the mean transverse momentum broadening squared per
unit length and is proportional to gluon distribution density
inside the nucleus. Here $\rho^A_N(\xi_N)$ is the single nucleon
density inside the nucleus and $p^+$ is the longitudinal momentum
per nucleon. Inclusion of higher-twist nucleon gluon
matrix elements under an extended adjoint two-gluon correlation
approximaton will give rise to a similar nuclear transverse
momentum distribution with an effective transverse-scale-dependent
($\xi_\perp^2=-\nabla_{k_\perp}^2$) transport parameter,
\begin{eqnarray}
\hat q_F(\xi_N, \xi_\perp^2)
&=&\frac{2\pi^2\alpha_s}{N_c}\rho^A_N(\xi_N)
\int \frac{d\xi^-}{2\pi p^+}(-2)
\langle N \mid
{\rm Tr} \left[
F_{+\sigma}(0,\vec 0_\perp){\cal L}_{\rm TMD}(0,\xi)
F_+^{\,\,\sigma}(\xi^-,\vec \xi_\perp)
{\cal L}_{\rm TMD}^\dagger(0,\xi)\right]
\mid N \rangle,
\end{eqnarray}
which is related to the transverse-scale-dependent
gluon distribution function in a nucleon. Such transverse-scale-dependent
transport parameter constituents power corrections to the
Gaussian form of the final nuclear broadening distribution.

The rest of this paper is organized as follows. In the next
section, we will first formulate the nuclear TMD quark
distribution in terms of collinear nuclear high-twist parton
matrix elements in the light-cone gauge and then derive the
nuclear transverse momentum broadening distribution under the
maximal two-gluon correlation approximation. We then include the
higher-twist nucleon quark matrix elements to consider the effect
of nucleon intrinsic transverse momentum. The derivation of
nuclear broadening of the TMD quark distribution is generalized to
any arbitrary gauge with the final results expressed in an
explicit gauge invariant form. Effects of higher-twist gluon
distribution functions are also discussed and are shown to
lead to a transverse-scale-dependent quark transport parameter. In
Sec.~\ref{sec:thermal}, we will extend the results on nuclear
transverse momentum broadening to the case of quark propagation in
hot medium and compare our results under the maximal two-gluon
approximation to the dipole model approximation in the Wilson line
approach to multiple parton scattering. We will also compare to
the result from ${\cal N}=4$ super-symmetric Yang-Mills (SYM)
calculation \cite{Liu:2006ug} of the transverse momentum
broadening and discuss the importance of multiple gluon
correlation in a strongly coupled system. Finally we will give a
summary in Sec.~\ref{sec:summary}.

%%%%%%%%%%%%%%%%%%%%%%%%%%%%%%%%%%%%%%%%%%%%%%%%%%%%%%%%%%%%%%%%%%%%%
%%%%%%%%%%%%%%%%%%%%%%%%%%%%%%%%%%%%%%%%%%%%%%%%%%%

\section {Nuclear transverse momentum broadening}
\label{sec-nutmd}

We will consider DIS off a large nucleus as depicted in Fig.~\ref{fig1}.
In the infinite momentum frame, the nucleus has a longitudinal
momentum $p=[p^+,0,\vec 0_\perp]$ per nucleon and a quark with fractional
longitudinal momentum $x_B=Q^2/2p^+q^-$ is knocked out by a virtual photon
with four-momentum $q=[-Q^2/2q^-, q^-, \vec 0_\perp]$. The differential cross
section for $e^-(l_1)+A (Ap) \to e^-(l_2)+q(k)+X$ can be written as,
\begin{equation}
E_{l_2}E_{k}\frac{d\sigma}{d^3l_2d^3k}
=\frac{\alpha_{\rm EM}^2}{2\pi s}\frac{1}{Q^4}
L^{\mu\nu}(l_1,l_2)E_{k}\frac{dW_{\mu\nu}}{d^3k}
\end{equation}
where $s=(p+l_1)^2$ and $alpha_{rm EM}$ is the fine structure constant in electrodynamics.
The SIDIS hadronic tensor $W_{\mu\nu}$ is defined as,
\begin{equation}
E_{k}\frac{dW_{\mu\nu}}{d^3k}= \sum_X
\langle A| J_\mu(0)|k,X\rangle \langle k,X| J_\nu(0)|A\rangle
2\pi\delta^4(p+q-k-p_X),
\end{equation}
and $J_\mu(y)=\bar\psi(y)\gamma_\mu \psi(y)$ is the hadronic electromagnetic
current. The leptonic tensor $L^{\mu\nu}$
is defined as usual and is given by,
\begin{equation}
L^{\mu\nu}(l_1,l_2)=\frac{1}{2}
{\rm Tr}[l_1\!\!\!/ \gamma_\mu l_2\!\!\!/ \gamma_\mu ]
=2[l_1^\mu{l_2}^\nu+l_1^\nu{l_2}^\mu-(l_1\cdot l_2)g^{\mu\nu}].
\end{equation}

The struck quark carrying large negative longitudinal momentum
$q^-$ will suffer multiple soft scattering with the rest of the
nucleus before hadronization into hadrons. We assume that the
virtuality of the photon $Q^2$ is very large and consider now only
the lowest order of the hard partonic part. The soft interaction
as shown in Fig.~\ref{fig1} can be resummed and the final leading
twist SIDIS tensor \cite{Liang:2006wp}
\begin{equation}
\frac{dW_{\mu\nu}}{d^2k_\perp} =H^{(0)}_{\mu\nu}(x_Bp,q)
f_q^A(x_B,\vec k_\perp),
\end{equation}
can be factorized as the product of the lowest hard partonic part
$H^{(0)}_{\mu\nu}(x_Bp,q)$,
\begin{equation}
H_{\mu\nu}^{(0)}(x_Bp,q) =\frac{e^2_q}{2}
{\rm Tr}[p\!\!\!/ \gamma_\mu (x_Bp\!\!\!/+q\!\!\!/)\gamma_\nu ]
\frac{2\pi}{2p\cdot q}.
\end{equation}
and the nuclear TMD quark distribution
function $f_q^A(x_B,\vec k_\perp)$ as defined in Eq.~(\ref{tmd0}).
Since the final state interaction are already included in the nuclear
TMD quark distribution function as the gauge links, one should
be able to derive the nuclear transverse momentum broadening.

\subsection{Nuclear TMD quark distribution function in light-cone gauge}

One important feature of the complete and gauge invariant
nuclear TMD quark distribution function in Eq.~(\ref{tmd0}) is
the transverse gauge link ${\cal L}_\perp(-\infty;\vec y_\perp,\vec 0_\perp)$
which depends on the transverse gauge 
potential $\vec A_\perp(-\infty,\vec\xi_\perp)$ at $-\infty$ along 
the light-cone. Without it, the gauge links in
the TMD quark distribution would completely vanish in the light-cone gauge
and one will be misled to assume that the final state interactions
become absent.
Furthermore the TMD quark distribution without the transverse gauge link is
no longer gauge invariant under residual gauge transformation since
the transverse gauge potential at infinity does not vanish in the
light-cone gauge \cite{Ji:2002aa} and is closely related to the singularity
of the gluon propagator in the light-cone gauge which has to be properly
regularized. Therefore, the effects of final state interaction are
actually encoded in the transverse gauge link in the light-cone gauge
and cannot be casually discarded. For this reason, we will first derive 
the nuclear transverse mometnum broadening in light-cone gauge and
repeat the derivation later in an arbitrary gauge.

Let us first consider the nuclear TMD quark distribution function
in the light-cone gauge $A_+=0$. In this gauge all the
longitudinal gauge links vanish and we are left with only the
transverse gauge link in the TMD quark distribution function. We
first insert a $\delta$ function into the TMD quark distribution
function,
\begin{equation}
f_q^A(x,\vec k_\perp)=
\int d^2\ell_\perp f_q^A(x,\vec \ell_\perp)
\delta^{(2)}(\vec k_\perp -\vec\ell_\perp).
\end{equation}
Using a Taylor expansion of the $\delta$-function,
\begin{equation}
\delta^{(2)}(\vec k_\perp-\vec \ell_\perp)
=e^{-\vec \ell_\perp\cdot \vec\nabla_{ k_\perp}}\delta^{(2)}(\vec k_\perp),
\end{equation}
the quark transverse momentum distribution can be written as
\begin{eqnarray}
f_q^A(x,\vec k_\perp)&=&\int \frac{dy^-}{2\pi} \frac{d^2y_\perp}{(2\pi)^2}
d^2 \ell_\perp e^{ixp^+y^- -i\vec  \ell_\perp\cdot \vec y_\perp}
\nonumber \\
&&\hspace{0.5in}\times
\langle A \mid \bar\psi(0,\vec 0_\perp)
\frac{\gamma^+}{2}{\cal L}^\dagger_\perp(-\infty;\vec y_\perp,\vec 0_\perp)
\psi(y^-,\vec y_\perp)\mid A \rangle
e^{-\vec  \ell_\perp\cdot \vec\nabla_{ k_\perp}}\delta^{(2)}(\vec k_\perp) \nonumber \\
&&\hspace{-0.5in}
=\int \frac{dy^-}{2\pi} \frac{d^2y_\perp}{(2\pi)^2}d^2 \ell_\perp
e^{ixp^+y^- -i\vec  \ell_\perp\cdot \vec y_\perp}
\langle A \mid \bar\psi(0,\vec 0_\perp)\frac{\gamma^+}{2}
%\nonumber \\
%&&\hspace{0.5in}\times
e^{i\vec\partial_{y_\perp} \cdot \vec\nabla_{ k_\perp}}
{\cal L}^\dagger_\perp(-\infty;\vec y_\perp,\vec 0_\perp)
\psi(y^-,\vec y_\perp)\mid A \rangle
\delta^{(2)}(\vec k_\perp),
\label{tmd1}
\end{eqnarray}
after partial integration in the transverse coordinate $\vec y_\perp$. Since
both the quark field and the transverse gauge link
depend on $\vec y_\perp$, we
have
\begin{eqnarray}
i\vec\partial_{y_\perp}{\cal L}^\dagger_\perp(-\infty;\vec y_\perp,\vec 0_\perp)
&=&{\cal L}^\dagger_\perp(-\infty;\vec y_\perp,\vec 0_\perp)
[-g\vec A_\perp(-\infty,\vec y_\perp)
+i\vec\partial_{y_\perp}] \nonumber \\
%&=&{\cal L}^\dagger_\perp(-\infty;\vec y_\perp,\vec 0_\perp)
%\left[i\vec D_\perp(0,\vec y_\perp)+g\int_{-\infty}^{0} d\xi^-
%\vec F_{+\perp}(\xi^-,\vec y_\perp)\right]
%\nonumber \\
&=&{\cal L}^\dagger_\perp(-\infty;\vec y_\perp,\vec 0_\perp)
\left[i\vec D_\perp(y^-,\vec y_\perp)+g\int_{-\infty}^{y^-} d\xi^-
\vec F_{+\perp}(\xi^-,\vec y_\perp)\right],
\label{lgt-trans}
\end{eqnarray}
where $\vec D_\perp(y^-,\vec y_\perp)=\vec\partial_{y_\perp}
+ig\vec A_\perp(y^-,\vec y_\perp)$ is the covariant derivative.
In the last step of the above
equation we used the following identity,
\begin{eqnarray}
\vec A_\bot(-\infty^-,y_\bot)
&=&\vec A_\bot(y^-,y_\bot)
-\int_{-\infty}^{y^-} d\xi^-\partial_+\vec A_\bot(\xi^-,y_\bot),
%\nonumber\\
%&=&\vec A_\bot(0^-,y_\bot)
%-\int d\xi^- \theta(-\xi^-) \partial_+\vec A_\bot(\xi^-,y_\bot)
\end{eqnarray}
and $\partial_+\vec A_\perp=\vec F_{+\perp}$ in the
light-cone gauge.
Completing the integration over the transverse momentum
$\vec  \ell_\perp$ in Eq.~(\ref{tmd1})
will now produce a $\delta$-function $\delta^{(2)}(\vec y_\perp)$
which will set transverse coordinate
$\vec y_\perp=\vec 0_\perp$ at which the transverse gauge link will
disappear. We have then,
\begin{eqnarray}
f_q^A(x,\vec k_\perp)&=&\int \frac{dy^-}{2\pi}e^{ixp^+y^-}
\langle A \mid \bar\psi(0)\frac{\gamma^+}{2}
e^{\vec W_\perp(y^-) \cdot\vec\nabla_{ k_\perp}}
\psi(y^-)\mid A \rangle
\delta^{(2)}(\vec k_\perp).
\label{tmd2}
\end{eqnarray}
Here we define the transport operator $\vec W_\perp(y^-,\vec y_\perp)$
in the light-cone gauge as
\begin{equation}
\vec W_\perp(y^-,\vec y_\perp)\equiv i\vec D_\perp(y^-,\vec y_\perp)
+g\int_{-\infty}^{y^-}d\xi^-\vec F_{+\perp}(\xi^-,\vec y_\perp).
\end{equation}
Note that in the light-cone gauge, the transport operator is translational
invariant, 
\begin{equation}
\vec W_\perp(y^-,\vec y_\perp)=\vec W_\perp(0,\vec y_\perp),
\end{equation}
along the light-cone.
For brevity in notation we will suppress the transverse
coordinates whenever they are set to zero in the field operators,
\begin{equation}
{\cal O}(y^-,\vec 0_\perp)\equiv {\cal O}(y^-).
\end{equation}

Eq.~(\ref{tmd2}) is a general result for the transverse momentum
quark distribution function in the light-cone gauge. In the case of a
large nucleus target, we can make further simplifications under
the assumption of a weakly bound nucleus. We first expand the
exponential factor in Eq.~(\ref{tmd2}) in power of the transport
operator $\vec W_\perp(0)$. The expectation
value of any odd power of the operator under any unpolarized
nuclear state should vanish under the parity invariance. We therefore
are left only with the even-power terms of the expansion,
\begin{equation}
{\cal M}_{2n}\equiv\frac{1}{(2n)!}
\langle A \mid \bar\psi(0)\frac{\gamma^+}{2}
\left[\vec W_\perp(y^-) \cdot\vec\nabla_{ k_\perp}\right]^{2n}
\psi(y^-)\mid A \rangle .
\end{equation}

We will neglect the covariant derivative first in the transport
operator and consider first the twist-four nuclear matrix elements ($n=1$),
\begin{equation}
{\cal M}_2=\frac{g^2}{2}
\int dy^- e^{ixp^+y^-}
\int_{-\infty}^{y^-} d\xi^-_1 \int _{-\infty}^{y^-}d\xi^-_2
\langle A \mid \bar\psi(0)\frac{\gamma^+}{2}
F_{+i}(\xi^-_1)F_{+j}(\xi^-_2)
\psi(y^-)\mid A \rangle .
\end{equation}
Because a nucleus consists of nucleons which
are color singlet states, the quark and gluon fields could either
be all attached to a single nucleon or to two separate nucleons.
In the first case, all four parton fields in the above
correlation matrix elements are confined
to the size of a nucleon $y^-, \xi_1^-, \xi_2^- \sim r_N$. On the other hand,
if quark and gluon fields are confined to two separate nucleons,
$y^-, |\xi^-|=|\xi_1^- -\xi_2^-| \sim r_N$, the overall position of the gluon
field $\xi_N^-=(\xi_1^- +\xi_2^-)/2$ will follow the second nucleon and
are only confined to the size of the nucleus $R_A$. Therefore,
the quark-gluon correlation function in this
case will have a nuclear enhancement of the order $R_A/r_N\sim A^{1/3}$ as
compared to the first case where both quark and gluon fields are confined
to a single nucleon. As a two-parton correlation approximation for a
large nucleus target we will only keep the matrix elements with the
nuclear enhancement. We will also neglect the correlation between
different nucleons and assume the large nucleus as a weakly bound. 
The leading contribution to the above quark-gluon correlation function
will be then,
\begin{eqnarray}
{\cal M}_2&\approx&A\int dy^- e^{ixp^+y^-}
\langle N \mid \bar\psi(0)\frac{\gamma^+}{2}
\psi(y^-)\mid N \rangle \frac{g^2}{2}
\int_{-\infty}^{0} d\xi^-_1 \int_{-\infty}^{0}d\xi^-_2
\frac{1}{N_c}\langle\!\langle {\rm Tr}
\left[F_{+i}(\xi^-_1)F_{+j}(\xi^-_2)\right]
\rangle\!\rangle_A.
\end{eqnarray}
If we further assume the large and weakly bound nucleus
as a homogenous system of nucleons,
\begin{equation}
\langle\!\langle {\rm Tr}
\left[F_{+i}(\xi^-_1)F_{+j}(\xi^-_2)\right]
\rangle\!\rangle_A
=\langle\!\langle {\rm Tr}
\left[F_{+i}(0)F_{+j}(\xi^-_2-\xi_1^-)\right]
\rangle\!\rangle_A
\end{equation}
and the nuclear length is much larger than nucleon size due
to confinement, $|\xi^-| \ll \xi_N^-$, we can approximate
the quark-gluon correlation as \cite{Osborne:2002st,jorge},
\begin{eqnarray}
{\cal M}_2&\approx&Af_q^N(x)
\frac{-g^2}{2}\int_{-\infty}^{0} d\xi^-_N 
\rho_N^A(\xi_N)\int\frac{d\xi^-}{2p^+}
\langle N \mid F_{+\sigma}(0)F_+^{\sigma}(\xi^-)
\mid N \rangle\frac{\delta_{ij}}{2}\frac{1}{2N_c} \nonumber \\
&= &A f_q^N(x)\frac{\delta_{ij}}{4}\int d\xi_N^- \hat q_F(\xi_N),
\end{eqnarray}
where
\begin{equation}
\langle\!\langle\cdots\rangle\!\rangle_A
=\int \frac{d^3p_N}{(2\pi)^3 2p^-}f_A(p_N,\xi_N)
\langle N \mid \cdots \mid N \rangle
=\frac{1}{2 p^+} \rho^A_N(\xi_N)\langle N \mid \cdots \mid N \rangle,
\end{equation}
denotes the ensemble (medium) average and $\rho_N^A(\xi_N)$ is the spatial
nucleon density inside the nucleus normalized to the atomic number $A$. 
The quark transport
parameter $\hat q_F(\xi_N)$ is defined as
\begin{equation}
\hat q_F(\xi_N)=-\frac{g^2}{2N_c}\rho_N^A(\xi_N)\int \frac{d\xi^-}{2p^+}
\langle N \mid F_{+\sigma}(0)F_+^{\sigma}(\xi^-)
\mid N \rangle
=\frac{2\pi^2\alpha_s}{N_c}\rho_N^A(\xi_N)[xf_N^g(x)]_{x=0},
\label{qhat1}
\end{equation}
and the gluon distribution function in a nucleon is
\begin{equation}
xf^N_g(x)=-\int_{-\infty}^{\infty}
\frac{d\xi^-}{2\pi p^+} e^{ixp^+\xi^-}
\langle N \mid F_{+\sigma}(0)F_+^{\;\;\sigma}(\xi^-)
\mid N \rangle ,
\end{equation}
where the summation over the gluon's color index in the matrix element
is implied in the definition of the gluon distribution function.

The TMD quark distribution function in Eq.~(\ref{tmd0}) is a leading-twist
result in terms of the momentum scale ($Q^2$) dependence of the DIS
process. Among higher-twist corrections, one has neglected those
from the transverse phase factors such as
\begin{equation}
e^{ix_\perp p^+(\xi_1^--\xi_2^-)};\,\, 
x_\perp=\frac{k_\perp^2}{2p^+q^-}=x_B\frac{k_\perp^2}{Q^2},
\end{equation}
that the propagating quark accumulates in the above two-gluon correlation 
matrix element.  They generally lead to contributions that are proportional
to higher-twist nuclear parton matrix elements and are power suppressed
${\cal O}(1/Q^{2n}), n\ge 1$. One can resurrect these higher-twist contributions
by substituting $k_\perp^2$ with its average value and setting the 
fractional momentum $x=x_\perp$ in the nucleon gluon distribution function 
in the quark transport parameter in Eq.~(\ref{qhat1}). This will introduce
the $Q^2$ or energy dependence of the quark transport parameter \cite{jorge}.
For the rest of this paper we will focus our attention to the leading-twist
TMD nuclear quark distribution.

In the above approximation of the twist-four nuclear
quark-gluon matrix we have neglected
multiple-nucleon correlation in a large nucleus. Such an approximation is
violated for small $x$ where quark-gluon and gluon-gluon fusion from different
nucleons become important and can lead to modification of the quark
distribution function and gluon saturation in a large
nucleus \cite{Gribov:1984tu,Mueller:1985wy,Iancu:2003xm}.
We also neglected the real part of the nucleon gluon matrix
elements which is responsible for Pomeron-like elastic (or
diffractive) scattering
and the nuclear shadowing of the quark and gluon distribution
functions \cite{Brodsky:1989qz,Nikolaev:1990ja,Huang:1997ii,Frankfurt:2002kd}
One can effectivly take into account these effects
by using a nuclear modified quark distribution function
$f_q^A(x_B)\neq A f_q^N(x_B)$ and
saturated gluon distribution function in the transport
parameter $\hat q_F$ which could lead a non-trivial nuclear and energy
dependence \cite{jorge}.

For other higher-twist nuclear matrix elements, we similarly
consider only the case where quark and gluon fields are attached to
different nucleons inside the nucleus.
\begin{eqnarray}
{\cal M}_{2n}
\approx Af_q^N(x)
\frac{1}{(2n)! N_c}\langle\!\langle{\rm Tr} \left[\vec W_\perp(y^-)
\cdot\vec\nabla_{ k_\perp}\right]^{2n}
\rangle\!\rangle_A = Af_q^N(x)
\frac{1}{(2n)! N_c}\langle\!\langle{\rm Tr} \left[\vec W_\perp(0)
\cdot\vec\nabla_{ k_\perp}\right]^{2n}
\rangle\!\rangle_A.
\label{maxcorr1}
\end{eqnarray}
Again we will only keep the
dominant terms that have the maximum nuclear enhancement for each
given power $2n$ (or twist) of the transport operator. Such contributions
come from contracting one pair of the gluonic fields with one
nucleon inside the large nucleus. Because of color confinement,
the relative longitudinal coordinate of the gluon pair is limited to
the size of the nucleon while the average coordinate is set by the
position of the nucleon which can be anywhere inside the nucleus.
Therefore, each pair of the gluon fields will give rise to one power
of nuclear enhancement factor $R_A\sim A^{1/3}$.

We will also neglect all terms that contain any power of the covariant
derivative $\vec D_\perp(0)$ in $\vec W_\perp(0)$ since they are
subleading in the nuclear enhancement comparing to the same twist
nuclear matrix elements without any covariant derivatives.
We call the above approximation {\em maximal
two-gluon correlation approximation} since we reduce the multiple gluon
correlations in the nucleus to  products of two-gluon
correlations that have the maximum nuclear size enhancement.
The leading contribution to the $2n$ gluon correlation
function is then
\begin{eqnarray}
\frac{1}{N_c}\langle\!\langle {\rm Tr} \left[\vec W_\perp(y^-)
\cdot\vec\nabla_{ k_\perp}\right]^{2n}
\rangle\!\rangle_A \approx \frac{(2n)!}{2^n n!}
\left[\frac{g^2}{2N_c}\frac{-1}{2p^+}
\int d\xi^-_N \rho^A_N(\xi_N) d\xi^-  \langle N \mid F_{+\sigma}(0)F_+^\sigma(\xi^-)
\mid N \rangle\frac{\nabla_{ k_\perp}^2}{2}\right]^n.
\label{2nfact}
\end{eqnarray}
This is essentially the extension of the approximation for
twist-four quark-gluon correlation in a large nucleus to the case
of quark-$n$-gluon correlation in which we assume the correlation
of $2n$ gluon fields is approximately the product of $n$ two-gluon
correlators (or gluon distribution functions). In the above
equation, the combinatorial factor for grouping $2n$ number of
gluon field operators into $n$ pairs,
\[ (2n-1)!! =\frac{(2n)!}{2^n n!}, \]
and the color factor for $2n$ gluon insertions,
\[
\frac{1}{N_c(N_c^2-1)^n}{\rm Tr}\left[T^{a_1}\cdots T^{a_n}T^{a_n}\cdots T^{a_1}\right]
=\frac{C_F^n}{(N_c^2-1)^n}=\frac{1}{(2N_c)^n},
\]
are used. Summation over polarization and color indices in the matrix
elements for quark and gluon distributions are implied. For gluon
propagation, the above color factor should be $C_A^n/(N_c^2-1)^n$ instead.

Using the definition of the quark transport parameter in nuclear
matter $\hat q_F(\xi_N)$ as defined in Eq.~(\ref{qhat1}), we can
now express the power expansion of the matrix elements as
\begin{eqnarray}
{\cal M}_{2n}&\approx&\frac{1}{(2n)!}
\int dy^- e^{ixp^+y^-}\langle A \mid \bar\psi(0)\frac{\gamma^+}{2}
\left[g\int_{-\infty}^{y^-} d\xi^-
\vec F_{+\perp}(\xi^-)\cdot\vec\nabla_{ k_\perp}\right]^{2n}
\psi(y^-)\mid A \rangle  \nonumber \\
&\approx&A f_q^N(x)
\frac{1}{n!}\left[\int d\xi^-_N \hat q_F(\xi_N)
\frac{\nabla_{ k_\perp}^2}{4}\right]^n\; ,
\end{eqnarray}

With the above simplification of the dominant contributions to
the nuclear matrix elements, we obtain the transverse momentum
distribution of the struck quark in DIS off a large nucleus
from Eq.~(\ref{tmd2}),
\begin{equation}
f_q^A(x,\vec k_\perp)\approx Af_q^N(x)
\exp\left[\int d\xi^-_N \hat q_F(\xi_N)
\frac{\nabla_{ k_\perp}^2}{4}\right] \delta^{(2)}(\vec  k_\perp)\,
\label{tmd3}
\end{equation}
in terms of collinear factorized (or transverse momentum integrated)
quark distribution functions and the quark transport parameter $\hat q_F$
which in turn is related to the gluon distribution density inside the nucleus.
This result is also recently obtained by Majumder and
M\"uller \cite{Majumder:2007hx} via direct resummation of all
twist diagrams in a covariant gauge calculation.

From Eq.~(\ref{tmd3}) one can then calculate the total transverse
momentum broadening of the struck quark due to multiple scattering
inside the nuclear matter,
\begin{equation}
\Delta_{2F}=
\frac{1}{Af_q^N(x)}
\int d^2 k_\perp  k_\perp^2 f_q^A(x,\vec k_\perp)
=\int d\xi^-_N \hat q_F(\xi_N)
\label{totbrd},
\end{equation}
which is the same as the twist-4 contribution \cite{lwz07}.
Though other multiple parton scatterings contribute to the
modified transverse momentum distribution, they do not affect
the broadening of the mean transverse momentum squared within
the maximal two-gluon correlation approximation.

If we define transverse (coordinate) distribution as
\begin{equation}
f_q^A(x,\vec y_\perp)=\int d^2k_\perp e^{i\vec k_\perp\cdot\vec y_\perp}
f_q^A(x,\vec k_\perp),
\end{equation}
The corresponding nuclear quark transverse distribution is,
\begin{eqnarray}
f_q^A(x,\vec y_\perp)&=&\int \frac{dy^-}{2\pi}e^{ixp^+y^-}
\langle A \mid \bar\psi(0)\frac{\gamma^+}{2}
e^{-i\vec y_\perp\cdot\vec W_\perp(y^-)}
\psi(y^-)\mid A \rangle
\approx Af_q^N(x) \langle\!\langle 
e^{-i\vec y_\perp\cdot\vec W_\perp(y^-)} \rangle\!\rangle_A
\nonumber \\
&\approx&
\exp\left[-\int d\xi^-_N \hat q_F(\xi_N)
\frac{y_\perp^2}{4}\right]
 Af_q^N(x),
\label{cotmd0}
\end{eqnarray}
which has a Gaussian form in the transverse coordinate $\vec y_\perp$. It
is then easy to obtain nuclear TMD quark distribution function as
\begin{equation}
f_q^A(x,\vec k_\perp)\approx Af_q^N(x)
\frac{1}{\pi\Delta_{2F}}\exp\left[- k_\perp^2/\Delta_{2F}\right],
\end{equation}
which is again a Gaussian with width given by the total transverse
momentum broadening squared $\Delta_{2F}$.

\subsection{Effect of nucleon TMD quark distribution}
\label{sec:ntmd}

In terms of twist expansion in the collinear factorization, one can
consider the nuclear modified transverse momentum distribution in
Eq.~(\ref{tmd3}) as the summation of all twist contributions. However,
it contains only contributions with the dominant nuclear enhancement
$A^{n/3}$ in the  $2(n+1)$-twist multiple-parton correlation
inside a large nucleus. Such dominant multi-parton correlations in
a large nucleus are shown to be made up of the products of leading
twist nucleon parton distributions. We have neglected higher-twist
contributions to the nucleon parton distribution, for example,
the intrinsic transverse momentum of quarks inside a nucleon.

Since the expression for the TMD nuclear parton distribution
function in Eq.~(\ref{tmd2}) is general, it should contain these
higher-twist effects. To isolate the contributions of the intrinsic
quark transverse momentum inside a nucleon, we will make the following
expansion of the matrix element in Eq.~(\ref{tmd2}),
\begin{eqnarray}
&&\langle A \mid \bar\psi(0)\frac{\gamma^+}{2}
e^{\vec W_\perp(y^-)\cdot\vec\nabla_{ k_\perp}}
\psi(y^-)\mid A \rangle
=\sum_{n=0}^{\infty}\frac{1}{(2n)!}
\langle A \mid \bar\psi(0)\frac{\gamma^+}{2}
\left[\vec W_\perp(y^-)\cdot\vec\nabla_{ k_\perp}\right]^{2n}
\psi(y^-)\mid A \rangle \nonumber \\
&&\hspace{0.5in} \approx\sum_{n=0}^{\infty}\frac{1}{(2n)!}\sum_{m=0}^{n}
\frac{(2n)!}{(2m)!(2n-2m)!} \frac{1}{N_c} \langle\!\langle
{\rm Tr}\left[\vec W_\perp(y^-)\cdot\vec\nabla_{ k_\perp}\right]^{2m}
\rangle\!\rangle_A
\nonumber \\
&&\hspace{1.0in} \times A\langle N \mid \bar\psi(0)\frac{\gamma^+}{2}
\left[\vec W_\perp(y^-)\cdot\vec\nabla_{ k_\perp}\right]^{2n-2m}
\psi(y^-)\mid N \rangle
\nonumber \\
&&\hspace{0.5in}\approx \sum_{m=0}^{\infty}\frac{1}{m!}
\left[\frac{g^2}{2N_c}\frac{1}{2p^+}
\int d\xi^-_N d\xi \rho^A_N(\xi_N) \langle N \mid F_{+\sigma}(0)F_+^\sigma(\xi^-)
\mid N \rangle\frac{\nabla_{ k_\perp}^2}{4}\right]^m
\nonumber \\
&&\hspace{1.0in}\times \sum_{n=m}^{\infty}\frac{A}{(2n-2m)!}
\langle N \mid \bar\psi(0)\frac{\gamma^+}{2}
\left[\vec W_\perp(y^-)\cdot\vec\nabla_{ k_\perp}\right]^{2n-2m}
\psi(y^-)\mid N \rangle
\nonumber \\
&&\hspace{0.5in}=\exp\left[\int d\xi^-_N \hat q_F(\xi_N)
\frac{\nabla_{ k_\perp}^2}{4}\right]
A \langle N \mid \bar\psi(0)\frac{\gamma^+}{2}
e^{\vec W_\perp(y^-)\cdot\vec\nabla_{ k_\perp}} \psi(y^-)\mid N \rangle,
\end{eqnarray}
where the approximation for the expectation value of $m$ pair of
gluon fields in a nucleus in Eq.~(\ref{2nfact}) is used.
Substitute the above matrix element into Eq.~(\ref{tmd2}), we
obtain the nuclear TMD parton distribution function,
\begin{equation}
f_q^A(x,\vec k_\perp)=A \exp\left[\int d\xi^-_N \hat q_F(\xi_N)
\frac{\nabla_{ k_\perp}^2}{4}\right] f_q^N(x,\vec k_\perp),
\label{tmd4}
\end{equation}
which is now related to the TMD parton distribution function of
the nucleon, $f_q^N(x,\vec k_\perp)$. Comparing to the
Eq.~(\ref{tmd3}), in which the nucleon TMD quark distribution
is assumed to be just a $\delta$-function, the nucleon intrinsic
transverse momentum in the above equation is the result of the
inclusion of a subset of non-leading (in nuclear enhancement)
higher-twist contributions.

From the above nuclear TMD quark distribution one
can derive a diffusion equation \cite{Majumder:2007hx,Baier:1996sk} for
the evolution of the quark transverse momentum distribution with
the nuclear size (or propagation length),
\begin{equation}
\frac{\partial f_q^A(x,\vec k_\perp)}{\partial\xi^-_N}
=\frac{1}{4}\hat q_F(\xi_N) \nabla_{ k_\perp}^2
 f_q^A(x,\vec k_\perp) ,
\end{equation}
with the diffusion constant given by the quark transport
parameter $\hat q_F(\xi_N)$. Apparently, this is the reason why
$\hat q_F(\xi_N)$ is often referred to as quark transport
coefficient \cite{Baier:1996sk}.

In coordinate space, nuclear quark transverse distribution can be
obtained from Eqs.~(\ref{tmd2}) and (\ref{tmd4}) by partial
integration,
\begin{eqnarray}
f_q^A(x,\vec y_\perp)&=&\int \frac{dy^-}{2\pi}e^{ixp^+y^-}
\langle A \mid \bar\psi(0)\frac{\gamma^+}{2}
e^{-i\vec W_\perp(y^-) \cdot\vec y_\perp}
\psi(y^-)\mid A \rangle
\approx A f_q^N(x,\vec y_\perp)
\langle\!\langle 
e^{-i\vec y_\perp\cdot\vec W_\perp(y^-)} \rangle\!\rangle_A
\nonumber \\
&\approx&
\exp\left[-\int d\xi^-_N \hat q_F(\xi_N)
\frac{y_\perp^2}{4}\right]
 A f_q^N(x,\vec y_\perp),
\label{cotmd}
\end{eqnarray}
as the product of the nucleon transverse distribution and a
Gaussian. The final quark transverse momentum distribution can
then be obtained from the Fourier transform of the above,
\begin{equation}
f_q^A(x,\vec k_\perp)=\frac{A}{\pi \Delta_{2F}}
\int d^2\ell_\perp e^{-(\vec k_\perp -\vec\ell_\perp)^2/\Delta_{2F}}f_q^N(x,\vec\ell_\perp),
\label{tmd5}
\end{equation}
as a convolution of the nucleon TMD quark distribution and a
Gaussian with a width $\Delta_{2F}$ given by the path integral of the
quark transport parameter or the total transverse momentum
broadening squared [Eq.~(\ref{totbrd})]. This is also a solution
to the diffusion equation with an initial condition at $\xi^-_N=0$
given by the nucleon TMD quark distribution function,
$Af_q^N(x,\vec k_\perp)$.

\subsection{Arbitrary gauge}

In an arbitrary gauge, one has to include both the longitudinal
and transverse gauge links in the gauge invariant definition of
the TMD parton distribution function in Eq.~(\ref{tmd0}).
Following the same procedure as in the light-cone gauge, one can
rewrite the nuclear TMD quark distribution function as
\begin{eqnarray}
f_q^A(x,\vec k_\perp)&=&\int \frac{dy^-}{2\pi} \frac{d^2y_\perp}{(2\pi)^2}
d^2\ell_\perp e^{ixp^+y^- -i\vec \ell_\perp\cdot \vec y_\perp}
e^{-\vec \ell_\perp\cdot \vec\nabla_{ k_\perp}}\delta^{(2)}(\vec k_\perp)
\langle A \mid \bar\psi(0,\vec 0_\perp)
\frac{\gamma^+}{2}
{\cal L}_{\rm TMD}(0,y)
\psi(y^-,\vec y_\perp)\mid A \rangle
\nonumber \\
&&\hspace{-0.3in}=
\int \frac{dy^-}{2\pi} \frac{d^2y_\perp}{(2\pi)^2}d^2\ell_\perp
e^{ixp^+y^- -i\vec \ell_\perp\cdot \vec y_\perp}
\langle A \mid \bar\psi(0,\vec 0_\perp)\frac{\gamma^+}{2}
e^{i\vec\partial_{y_\perp} \cdot \vec\nabla_{ k_\perp}}
{\cal L}_{\rm TMD}(0,y)
\psi(y^-,\vec y_\perp)\mid A \rangle
\delta^{(2)}(\vec k_\perp).
\label{tmdco1}
\end{eqnarray}
The transverse differentiation should act on both the quark field
operator $\psi(y^-,\vec y_\perp)$ and the gauge
link ${\cal L}_{\rm TMD}(0,y)$. We will use the following
identity [Eq.~(\ref{lc-diff}) in Appendix A],
\begin{eqnarray}
i\vec\partial_{y_\perp}{\cal L}_{\rm TMD}(0,y)
={\cal L}_{\rm TMD}(0,y)
\vec W_\perp(y^-,\vec y_\perp)
\end{eqnarray}
with the gauge covariant form of the
transport operator $\vec W_\perp(y^-,\vec y_\perp)$ given by,
\begin{eqnarray}
\vec W_\perp(y^-,\vec y_\perp)&\equiv&
i\vec D_\perp(y^-,\vec y_\perp)
+g\int_{-\infty}^{y^-}d\xi^-
{\cal L}^\dagger_\parallel(\xi^-,y^-;\vec y_\perp)
\vec F_{+\perp}(\xi^-,\vec y_\perp)
{\cal L}_\parallel(\xi^-,y^-;\vec y_\perp),
\label{transop}
\end{eqnarray}
which transforms like a covariant derivative $\vec
D_\perp(y^-,\vec y_\perp)$ under any gauge transformation. One can
recover from the above the equivalent identity in the case of
light-cone gauge in Eq.~(\ref{lgt-trans}) by setting $A_+=0$.
These identities have been used to relate the T-odd and
spin-dependent part of the quark distribution function to twist-3
parton matrix elements \cite{Eguchi:2006mc,Boer:2003cm}. These
twist-3 matrix elements are related to the first moments in
transverse momentum of the TMD parton distribution function and
the twist-3 contribution to DIS process.

Completing the integration over the transverse momentum $\vec k_\perp$
and coordinate $\vec y_\perp$, we can recast the nuclear TMD parton
distribution function as,
\begin{eqnarray}
f_q^A(x,\vec k_\perp)&=&\int \frac{dy^-}{2\pi}e^{ixp^+y^-}
\langle A \mid \bar\psi(0)\frac{\gamma^+}{2}
{\cal L}_\parallel(0,y^-)
e^{\vec W_\perp(y^-) \cdot\vec\nabla_{ k_\perp}}
\psi(y^-)\mid A \rangle
\delta^{(2)}(\vec k_\perp).
\label{tmdco2}
\end{eqnarray}
It is in exactly the same form as the expression in the light-cone
gauge in Eq.~(\ref{tmd2}) except that there is
the longitudinal gauge link which
is necessary to ensure the gauge invariance of the above form of
nuclear TMD parton distribution function. The transport
operator $\vec W_\perp(y^-)\equiv \vec W_\perp(y^-,\vec 0_\perp)$
is now given by its more general form
in Eq.~(\ref{transop}). Integrating over
the transverse momentum, one obtains the collinear
factorized (or transverse momentum integrated) quark distribution
function,
\begin{equation}
f_q^A(x)=\int \frac{dy^-}{2\pi}e^{ixp^+y^-}
\langle A \mid \bar\psi(0)\frac{\gamma^+}{2}
{\cal L}_\parallel(0,y^-)
\psi(y^-)\mid A \rangle ,
\end{equation}
that is also gauge invariant under any arbitrary gauge
transformation. In the explicit calculation of the transverse
momentum broadening via cut-diagrams as in
Ref.~\cite{Majumder:2007hx}, these gauge links arise from
resummation of extra number of collinear gluons on either side of
the cut in addition to the soft gluons with transverse momentum
that contribute to the transverse momentum of the final quark.

Following the same steps as in the case of light-cone gauge, we
will be able to derive from Eq.~(\ref{tmdco2}) the nuclear
modified transverse momentum distribution function in an arbitrary
gauge as given in Eqs.~(\ref{tmd4}) and (\ref{tmd5}). The
corresponding quark transport parameter $\hat q_F$ is simply
replaced by a gauge invariant form
\begin{eqnarray}
\hat q_F(\xi_N)&=&\frac{2\pi^2\alpha_s}{N_c}\rho^A_N(\xi_N)
[xf_g^N(x)]_{x=0};
\nonumber \\
xf_g^N(x)&=&-2
\int \frac{d\xi^-}{2\pi p^+}
e^{ixp^+\xi^-}
\langle N \mid {\rm Tr} \left[
F_{+\sigma}(0){\cal L}_\parallel(0,\xi^-)F_+^\sigma(\xi^-)
{\cal L}_\parallel(\xi^-,0)\right]
\mid N \rangle,
\label{fnlqhat}
\end{eqnarray}
where the gluon field is expressed in the fundamental representation
$F_{+\sigma}=F_{+\sigma}^aT^a$. Note that the above definition of
the gauge invariant gluon distribution function in the fundamental
color representation is equivalent to the definition in the
adjoint representation \cite{Collins:1981uk,Ji:2005nu},
\begin{equation}
xf_g^N(x)=-
\int \frac{d\xi^-}{2\pi p^+}
e^{ixp^+\xi^-}
\langle N \mid
F^a_{+\sigma}(0){\cal L}^A_{\parallel ab}(0,\xi^-)F_+^{b\,\sigma}(\xi^-)
\mid N \rangle,
\end{equation}
where
\begin{equation}
{\cal L}^A_{\parallel}(0,\xi^-)\equiv
\exp\left[-i g\int_{\xi^-}^0 d\zeta^- A_+^c(\zeta^-,\vec 0_\perp)t^c_A\right],
\end{equation}
with $(t_A^c)_{ab}=-if_{abc}$, is the longitudinal gauge link in the
adjoint representation. One can similarly introduce the
transverse gauge link in the adjoint representation
\begin{equation}
{\cal L}^A_{\perp}(-\infty;\vec \xi_\perp, \vec 0_\perp)
\equiv
\exp\left[-i g\int^{\vec \xi_\perp}_{\vec 0_\perp}
 d\vec\xi_\perp\cdot \vec A_\perp^c(-\infty,\vec\xi_\perp)t^c_A\right],
\end{equation}
and the TMD gluon distribution function,
\begin{eqnarray}
xf_g^N(x,\vec k_\perp)&=&-
\int \frac{d^2\xi_\perp}{(2\pi)^2} \frac{d\xi^-}{2\pi p^+}
e^{ixp^+\xi^- -i\vec \xi_\perp\cdot\vec k_\perp}
\langle N \mid
F^a_{+\sigma}(0,\vec 0_\perp)
{\cal L}^{A}_{{\rm TMD}ab}(0,\xi)
F_+^{b\,\sigma}(\xi^-,\vec\xi_\perp)
\mid N \rangle,
\nonumber \\
&=&-\int\frac{d\xi^-}{2\pi p^+}
e^{ixp^+\xi^-}
\langle N \mid
F^a_{+\sigma}(0)
\left[e^{\vec W_\perp^A(\xi^-)
\cdot\vec\nabla_{ k_\perp}}\right]_{ab}
F_+^{b\,\sigma}(\xi^-)
\mid N \rangle \delta^{(2)}(\vec k_\perp),
\end{eqnarray}
where
\begin{equation}
{\cal L}^{A}_{{\rm TMD}ab}(0,\xi)=
{\cal L}^{A\dagger}_{\parallel ac}(-\infty,\xi^-;\vec 0_\perp)
{\cal L}^{A\dagger}_{\perp cd}(-\infty;\vec \xi_\perp, \vec 0_\perp)
{\cal L}^{A}_{\parallel db}(-\infty,\xi^-;\vec \xi_\perp),
\end{equation}
and
\begin{eqnarray}
\vec W_{\perp ab}^A(\xi^-,\vec \xi_\perp)&\equiv&
i\vec D^A_{\perp ab}(\xi^-,\vec \xi_\perp)
+g\int_{-\infty}^{\xi^-}d\zeta^-
{\cal L}^{A\dagger}_{\parallel ac}(\zeta^-,\xi^-;\vec \xi_\perp)
\vec F^A_{+\perp ce}(\zeta^-,\vec \xi_\perp)
{\cal L}^A_{\parallel eb}(\zeta^-,\xi^-;\vec \xi_\perp),
\label{cotransop}
\end{eqnarray}
is the transport operator in the adjoint representation. The covariant
derivative $D^A_{\perp}$ and gluon field strength 
$\vec F^A_{+\perp}$ in the adjoint representation are defined as
\begin{equation}
\vec D^A_{\perp ab}(\xi^-,\vec \xi_\perp)
=\delta_{ab}\vec \partial_\perp +
ig \vec A_\perp^c(\xi^-,\xi_\perp)(t^c_A)_{ab},\,\,\,\, 
\vec F^A_{+\perp ab}\equiv \vec F^c_{+\perp}(t^c_A)_{ab},
\end{equation}
respectively.
The corresponding transverse coordinate gluon distribution is then
\begin{eqnarray}
xf_g^N(x,\vec y_\perp)=-\int\frac{d\xi^-}{2\pi p^+}
e^{ixp^+\xi^-}
\langle N \mid
F^a_{+\sigma}(0)
\left[e^{-i \vec y_\perp \cdot \vec W_\perp^A(\xi^-)}\right]_{ab}
F_+^{b\,\sigma}(\xi^-) \mid N \rangle.
\label{tddgluon}
\end{eqnarray}

\subsection{Effect of nucleon TMD  gluon distribution}
\label{sec:tmdgluon}

In the maximal two-gluon correlation approximation, we approximate
the higher-twist nuclear parton matrix elements with  products of
twist-two nucleon parton matrix elements [see Eqs.~(\ref{maxcorr1})
and (\ref{2nfact})]. Higher-twist nucleon quark matrix elements and 
quark-gluon correlations lead to intrinsic transverse momentum distribution
inside the nucleon. However, we have so far neglected contributions from
higher-twist nucleon gluon matrix elements that involve covariant 
derivative $\vec D_\perp$ or multi-gluon correlation within a nucleon.
These matrix elements have non-leading nuclear length dependence as compared
to the products of twist-two gluon distributions. In order to consider the 
effect of these higher-twist nucleon gluon matrix elements, we separate 
the covariant derivative from the transport operator in light-cone 
gauge (to simplify notations),
\begin{eqnarray}
\vec W_\perp(y^-) &=&
\vec {\cal F}_\perp(y^-)+i\vec D_\perp(y^-);
\nonumber \\
\vec {\cal F}_\perp(y^-)&\equiv&
g\int_{-\infty}^{y^-}d\xi^-
\vec F_{+\perp}(\xi^-).
\end{eqnarray}
Note that [see Eq.~(\ref{adjco}) for arbitrary gauge in the
Appendix]
\begin{equation}
D_{\perp i}(y^-)\vec {\cal F}_\perp(y^-)=
g\int_{-\infty}^{y^-}d\xi^-
D^A_{\perp i}(\xi^-)\vec F_{+\perp}(\xi^-) +
\vec {\cal F}_\perp(y^-)D_{\perp i}(y^-)
\equiv
D^A_{\perp i}\vec{\cal F}_\perp(y^-) +
\vec {\cal F}_\perp(y^-)D_{\perp i}(y^-),
\end{equation}
where $\vec D_\perp^A F=\vec\partial_\perp F+ig[\vec A_\perp,F]$
is the covariant derivative in the adjoint representation. Since
one can factor out the covariant derivatives of the quark field 
together with other higher-twist nucleon quark-gluon matrix
elements into the TMD nucleon quark distribution function,
one can effectively replace the covariant derivative $\vec D_\perp$
with its adjoint form $\vec D_\perp^A$ in the the Taylor expansion 
of the nuclear gluon matrix element
\begin{equation}
f(y_\perp^2)=\frac{1}{N_c}\langle\!\langle
{\rm Tr} e^{-i \vec y_\perp\cdot\vec W_\perp(y^-)}\rangle\!\rangle_A,
\end{equation}
which should be a function of $y_\perp^2$ because of the parity
invariance of the unpolarized nucleus state. We will again
neglect nucleon correlations and assume homogeneity in the nucleus.
However, we now relax the maximal two-gluon correlation
approximation to include higher-twist nucleon matrix elements that
contain covariant derivatives and multiple gluon correlations. They
are considered sub-leading in the nuclear length dependence in the above
Taylor expansion. We denote $f_n(y_\perp^2)$ as the $n$th term in
the Taylor expansion of the matrix elements. Therefore, the linear
term in $y_\perp^2$ is
\begin{eqnarray}
f_1(y_\perp^2)=
-\frac{1}{2} \frac{1}{N_c}\langle\!\langle
{\rm Tr}\left[\vec y_\perp\cdot \vec W_\perp(y^-)\right]^2
\rangle\!\rangle_A
=-\frac{y_\perp^2}{4} \frac{1}{2N_c}\langle\!\langle
\vec W_\perp^a(y^-)\cdot
\vec W_\perp^a(y^-)\rangle\!\rangle_A
&=&-\frac{y_\perp^2}{4}\Delta_{2F}.
\end{eqnarray}
Note that the medium averaged value of matrix elements linear
in ${\cal F}_\perp(y^-)$ should vanish.

For the quadratic term in $y_\perp^2$, we first separate
the gluon correlation into connected and disconnected parts,
\begin{eqnarray}
f_2(y_\perp^2)&=&\frac{1}{4!N_c} \langle\!\langle{\rm Tr}
\left[\vec y_\perp\cdot \vec W_\perp(y^-)\right]^4
\rangle\!\rangle_A
=\left(\frac{y_\perp^2}{4}\right)^2\frac{1}{2!N_c} 
\langle\!\langle{\rm Tr}
\left[\vec W_\perp(y^-)\right]^4 \rangle\!\rangle_A
\nonumber \\
&&=\left(\frac{y_\perp^2}{4}\right)^2\frac{1}{2!}
\left\{\Delta_{2F}^2+\frac{1}{N_c}\langle\!\langle{\rm Tr}
\left[\vec W_\perp(y^-)\right]^4
\rangle\!\rangle_{AC}\right\}
\equiv \left(\frac{y_\perp^2}{4}\right)^2
\left[\frac{1}{2!}\Delta_{2F}^2+\Delta_{4F}\right],
\end{eqnarray}
using the identity for generators of the fundamental representation,
\begin{equation}
T^aT^b=\frac{1}{2N_c}\delta_{ab}+
\frac{1}{2}d_{abc}T^c+\frac{i}{2}f_{abc}T^c,
\end{equation}
where the connected parts of the matrix elements 
$\langle\!\langle\cdots\rangle\!\rangle_{AC}$ exclude the singlet
contribution in the above color decomposition. We call
$\Delta_{4F}$ twist-four quark transport parameter which
contains all the twist-four nucleon gluon matrix elements
in the connected part of the nuclear gluon matrix elements.

In evaluating the connected parts of the nuclear gluon matrix
elements, we will now adopt what we call {\em extended two-gluon}
correlation approximation, in which we separate two gluon fields
out of the nuclear matrix elements,
\begin{eqnarray}
\Delta_{4F}&=&\frac{1}{2N_c}\langle\!\langle{\rm Tr}
\left[\vec W_\perp(y^-)\right]^4
\rangle\!\rangle_{AC}\approx 
\frac{2}{2N_c}\langle\!\langle{\rm Tr}
\left[\vec {\cal F}_\perp(y^-)\cdot W_\perp^2(y^-)
\vec {\cal F}_\perp(y^-)\right]
\rangle\!\rangle_{AC}
\nonumber \\
&=&\frac{1}{N_c}\langle\!\langle{\rm Tr}
\left\{\vec {\cal F}_\perp\cdot 
\left[{\cal F}_\perp^2+2\vec{\cal F}_\perp\cdot i\vec D_\perp^A
+(i\vec D_\perp^A\cdot\vec{\cal F}_\perp)+(iD_\perp^A)^2\right]
\vec {\cal F}_\perp \right\}
\rangle\!\rangle_{AC}
\nonumber \\
&=&\frac{1}{2N_c}\langle\!\langle
\vec {\cal F}_\perp^a\cdot\left[({\cal F}^{O})^2_{ab}+
2(i\vec D_\perp^A\cdot\vec{\cal F}^O_\perp)_{ab}
+(i\vec D_\perp^A\cdot\vec{\cal F}_\perp^O)_{ab}
+(iD_\perp^A)^2_{ab} \right]\vec {\cal F}_\perp^b
\rangle\!\rangle_A
\nonumber \\
&\equiv& \frac{1}{2N_c}\langle\!\langle
\vec {\cal F}_\perp^a\cdot
\left[W_\perp^O\right]^2_{ab} \vec {\cal F}_\perp^b\rangle\!\rangle_A
=\int \frac{d\xi_1^-d\xi_2^-}{2\pi p^+}
\rho_N^A(\xi_N) \frac{\pi g^2}{2N_c}
\langle N \mid \vec F_{+\perp}^a(\xi_1^-)
\cdot \left[W_\perp^O\right]^2_{ab}
\vec F_{+\perp}^b(\xi_2^-)\mid N \rangle .
\end{eqnarray}
Note that there are 2 pairs of gluons for the extended two-gluon
correlation approximation and we have excluded the disconncted 
contribution (singlet)
from the trace operation. We have defined the {\em octect} gluon 
field strength
\begin{equation}
\vec{\cal F}_{\perp ab}^O\equiv \frac{1}{2}
\vec{\cal F}_\perp^c(d_{cab}-if_{cab})
\end{equation}
and the corresponding transport operator
\begin{equation}
\vec W_{\perp ab}^O\equiv \vec{\cal F}_{\perp ab}^O
+i\vec D_{\perp ab}^A.
\end{equation}
Again we have dropped terms linear in ${\cal F}_\perp(y^-)$.

Similarly, one can also get the third term in the Taylor
expansion of the transverse distribution,
\begin{eqnarray}
f_3(y_\perp^2)&=&\frac{-1}{6!N_c} \langle\!\langle{\rm Tr}
\left[\vec y_\perp\cdot \vec W_\perp(y^-)\right]^6
\rangle\!\rangle_A
=-\left(\frac{y_\perp^2}{4}\right)^3\frac{1}{3!N_c} 
\langle\!\langle{\rm Tr}
\left[\vec W_\perp(y^-)\right]^6 \rangle\!\rangle_A
\nonumber \\
&=&-\left(\frac{y_\perp^2}{4}\right)^3\frac{1}{3!}
\left\{\Delta_{2F}^3+3\Delta_{2F}\frac{1}{N_c}\langle\!\langle{\rm Tr}
\left[\vec W_\perp(y^-)\right]^4
\rangle\!\rangle_{AC}+
\frac{1}{N_c}\langle\!\langle{\rm Tr}
\left[\vec W_\perp(y^-)\right]^6
\rangle\!\rangle_{AC}
\right\}
\nonumber \\
&\equiv& \left(\frac{y_\perp^2}{4}\right)^3
\left[\frac{1}{3!}\Delta_{2F}^3+\Delta_{2F}\Delta_{4F}
+\frac{1}{2!}\Delta_{6F}\right].
\end{eqnarray}
We again assume the extended two-gluon correlation 
approximation,
\begin{eqnarray}
\Delta_{6F}&\equiv&\frac{2}{3!N_c}\langle\!\langle{\rm Tr}
\left[\vec W_\perp(y^-)\right]^6
\rangle\!\rangle_{AC}
\approx \frac{6}{3!N_c}\langle\!\langle{\rm Tr}
\left[\vec {\cal F}_\perp(y^-)\cdot
\vec W_\perp^4(y^-) \vec{\cal F}_\perp(y^-)\right]
\rangle\!\rangle_{AC}
=\frac{1}{2N_c}\langle\!\langle
\vec {\cal F}_\perp^a\cdot
\left[\vec W_\perp^O \right]^4_{ab} \vec{\cal F}_\perp^b
\rangle\!\rangle_{A}
\nonumber \\
&=&\int \frac{d\xi_1^-d\xi_2^-}{2\pi p^+}
\rho_N^A(\xi_N) \frac{\pi g^2}{2N_c}
\langle N \mid \vec F_{+\perp}^a(\xi_1^-)
\cdot \left[W_\perp^O\right]^4_{ab}
\vec F_{+\perp}^b(\xi_2^-)\mid N \rangle .
\end{eqnarray}

Following the same procedure and the extended two-gluon
approximation, we can obtain other terms in the Taylor expansion
and the final form of the transverse expansion,
\begin{eqnarray}
f(y_\perp^2)&\approx&1-\frac{y_\perp^2}{4}\Delta_{2F}+
\left(\frac{y_\perp^2}{4}\right)^2
\left[\frac{1}{2!}\Delta_{2F}^2+\Delta_{4F}\right]
-\left(\frac{y_\perp^2}{4}\right)^3
\left[\frac{1}{3!}\Delta_{2F}^3+\Delta_{2F}\Delta_{4F}
+\frac{1}{2!}\Delta_{6F}\right]+{\cal O}(y_\perp^8)
\nonumber \\
&=&\exp\left\{-\frac{y_\perp^2}{4}\left[\Delta_{2F}
-\frac{y_\perp^2}{4}\Delta_{4F}
+\left(\frac{y_\perp^2}{4}\right)^2\frac{1}{2!}\Delta_{6F}
+\cdots\right]\right\}
\equiv \exp\left[-\frac{y_\perp^2}{4}\Delta_{F}(y_\perp^2)\right] .
\label{hgtwst-dis}
\end{eqnarray}

As one can observe, inclusion of higher-twist nucleon gluon
matrix elements in the medium averaged products of the
transport operator will give rise to a transverse-distance-dependent (TDD)
quark transport parameter,
\begin{eqnarray}
\Delta_{F}(y_\perp^2)&\equiv&
\int d\xi_N^- \hat q_{F}(\xi_N,y_\perp^2)
\nonumber \\
&=&\int \frac{d\xi_1^-d\xi_2^-}{2\pi p^+}
\rho_N^A(\xi_N) \frac{\pi g^2}{2N_c}
\sum_{n=0}^{\infty} \frac{(-1)^n}{n!}
\left(\frac{y_\perp^2}{4}\right)^n
\langle N \mid \vec F_{+\perp}^a(\xi_1^-)
\cdot\left[W_\perp^O\right]^{2n}_{ab}
\vec F_{+\perp}^b(\xi_2^-)\mid N \rangle,
\label{tddtranspt}
\end{eqnarray}
that depends on higher-twist nucleon gluon matrix elements.
Such TDD quark transport parameter effectively contribute to power
corrections to the nuclear transverse momentum distribution and
render it a non-Gaussian form, especially in the small
transverse momentum or large transverse coordinate region.
However, these contributions have
sub-leading nuclear length dependence as compared to those
associated with the leading-twist quark transport parameter $\Delta_{2F}$.
Since the leading twist (twist-2) transport parameter
is proportional to the collinear nucleon gluon distribution function, 
such TTD transport parameter might be related to the TTD gluon
distribution function.

It is helpful, therefore, to make a similar Taylor expansion of the
TDD gluon distribution function from Eq.~(\ref{tddgluon}),
\begin{eqnarray}
[xf_g^N(x,y_\perp^2)]_{x=0}
&=&\int \frac{d\xi^-}{2\pi p^+}
\langle N \mid \vec F_{+\perp}^a(0)\cdot
\left[e^{-i\vec y_\perp\cdot\vec W^A_\perp(\xi^-)}\right]_{ab}
\vec F_{+\perp}^b(\xi^-)\mid N \rangle
\nonumber \\
&=&\int \frac{d\xi^-}{2\pi p^+}
\sum_{n=0}^{\infty}\frac{(-1)^n}{(2n)!}
\langle N \mid \vec F_{+\perp}^a(0)\cdot
\left[\vec y_\perp\cdot\vec W^A_\perp(\xi^-)\right]^{2n}_{ab}
\vec F_{+\perp}^b(\xi^-)\mid N \rangle
\nonumber \\
&=&\int \frac{d\xi^-}{2\pi p^+}
\sum_{n=0}^{\infty}\frac{(-1)^n}{n!}
\left(\frac{y_\perp^2}{4}\right)^n
\langle N \mid \vec F_{+\perp}^a(0)\cdot
\left[W^A_\perp(\xi^-)\right]^{2n}_{ab}
\vec F_{+\perp}^b(\xi^-)\mid N \rangle.
\end{eqnarray}
Comparing the above Taylor expansion of the TDD gluon distribution
function to the TDD quark transport parameter in Eq.~(\ref{tddtranspt}),
one can indeed relate the two,
\begin{equation}
\hat q_{F}(\xi_N,y_\perp^2)\approx\rho_N^A(\xi_N) \frac{\pi g^2}{2N_c}
[xf_g^N(x,y_\perp^2)]_{x=0},
\label{tddq-gdis}
\end{equation}
if we approximate the octet gluon field strength with its
adjoint value
\begin{equation}
\vec {\cal F}^O_{\perp ab}=\frac{1}{2}
\vec {\cal F}^c_\perp(d_{abc}-if_{abc})\approx
-i\vec {\cal F}^c_\perp f_{abc}=\vec {\cal F}^A_{\perp ab}.
\end{equation}
We will refer to the above approximation and the extended two-gluon
correlation approximation together as {\em extended adjoint two-gluon}
correlation approximation. Under such approximation, we can resum
contributions associated with the higher-twist nucleon
gluon matrix elements to a TDD quark transport parameter which
is related to TDD gluon distribution function. We would like
to point out that under the extended adjoint two-gluon correlation
approximation, one actully only includes a subset of higher-twist
nucleon gluon matrix elements (see Appendix B). This is similar to the
approximation we made in order to fact out the nucleon TMD
quark distribution from the nuclear TMD quark distribution
in Sec.~\ref{sec:ntmd}.

Note that even though the higher-twist contributions we have considered
so far lead to a non-Gaussian
nuclear transverse momentum broadening, they do not contribute to the
averaged transverse momentum broadening squared which is
still given by the twist-2 transport parameter $\Delta_{2F}$. These
higher-twist nucleon gluon matrx elements only contribute to higher
moments of the transverse momentum broadening.

One should also keep in mind that we have only considered the
leading order in $\alpha_s$ of the hard scattering. Higher-order
contributions should also lead to leading-twist non-Gaussian
components of the transverse momentum distribution \cite{Ji:2004wu}.

Finally, in a finite nucleus, one has to take into account the
finite number of nucleons $A$ in the nucleus when factorizing
the nuclear parton matrix elements into products of nucleon
parton matrix elements. Such consideration will lead to
a quark distribution function in the coordinate 
space \cite{Osborne:2002st},
\begin{eqnarray}
f(y_\perp^2)&=&1-(A-1)\frac{y_\perp^2}{4}
\int \frac{d\xi_N}{A} \hat q(\xi_N,y_\perp^2)
+\frac{(A-1)(A-2)}{2!}\left[\frac{y_\perp^2}{4}
\int \frac{d\xi_N}{A} \hat q(\xi_N,y_\perp^2)\right]^2
+\cdots
\nonumber \\
&=&\left[1-\frac{y_\perp^2}{4}
\int \frac{d\xi_N}{A} \hat q(\xi_N,y_\perp^2)\right]^{A-1},
\end{eqnarray}
which can be approximated as that in Eq.~(\ref{hgtwst-dis}) 
for a large nucleus $A\gg 1$.

\section{Quark propagation in a thermal medium}
\label{sec:thermal}

We can generalize our study of nuclear transverse momentum broadening
to quark propagation in a hot medium such as the quark-gluon plasma produced
in high-energy heavy-ion collisions. In this case, the initial quark
production cross section is assumed to be factorized from the quark
propagation in medium.

The thermal medium can be considered as an interacting gas of
colored constituents with a local density $\rho_N^A(\xi_N)$, with
$N$ now referring to the color constituents. The correlation
length among these constituents are determined by the screening
scale $1/\mu$ which could be longer than the inter-constituent
distance as given by the temperature $1/T$ in the weak coupling
limit. Under such scenario, one can still apply the maximal
two-gluon correlation approximation to the medium averaged
multiple gluon matrix elements as we have used in the cold nuclear
medium.

With an initial condition of longitudinal momentum $p^+$ and zero
transverse momentum, the final quark transverse momentum
distribution can be easily read from Eq.~(\ref{tmd4}),
\begin{eqnarray}
f_q^A(x,\vec k_\perp)&=&f(\vec k_\perp)\delta(x-1)
\nonumber \\
f(\vec k_\perp)&=&\exp\left[\int d\xi_N^- \hat q_F(\xi_N)
\frac{\nabla_{k_\perp}^2}{4}\right] \delta^{(2)}(\vec k_\perp)
=\frac{1}{\pi\Delta_{2F}}\exp\left[-\frac{k_\perp^2}{\Delta_{2F}}\right],
\end{eqnarray}
or in terms of the transverse coordinate distribution
\begin{equation}
f(\vec y_\perp)=\exp\left[-
\int d\xi_N^-\hat q_F(\xi_N)\frac{y_\perp^2}{4}\right]
\label{ydistr}
\end{equation}
for the final quark. From the general form of the TMD quark
distribution function in Eq.~(\ref{tmdco1}), one can replace the
nuclear state with a quark with momentum $[p^+,\vec 0_\perp]$ and
the medium $A$. Averaging over the initial state of both quark and
the medium, one obtains the transverse coordinate distribution for
a propagating quark as given by the medium expectation value of a
pure gauge link,
\begin{equation}
f(\vec y_\perp)=\frac{1}{N_c}
\langle\!\langle{\rm Tr}\left[
{\cal L}^\dagger_\parallel(-\infty,\infty;\vec 0_\perp)
{\cal L}_\perp(-\infty;\vec 0_\perp,\vec y_\perp)
{\cal L}_\parallel(-\infty,\infty;\vec y_\perp)\right]
\rangle\!\rangle,
\end{equation}
where we have assumed the quark is produced at $\infty$
and propagates toward $-\infty$ along the light-cone
according to the convention used in this paper.

\subsection{Dipole model approximation}

In a covariant gauge (where the transverse gauge link becomes unity),
the above becomes the Wilson line formulation of multiple
scattering \cite{Wiedemann:2000za,Kovner:2003zj}. Under a
dipole model, the medium averaged Wilson line can be
approximated \cite{Kovner:2001vi} as
\begin{eqnarray}
\frac{1}{N_c}
\langle{\rm Tr}\left[
{\cal L}^\dagger_\parallel(-\infty,\infty;\vec 0_\perp)
{\cal L}_\parallel(-\infty,\infty;\vec y_\perp)\right]\rangle
\approx \exp\left[-\frac{1}{2}\int d\xi_N^-
\rho_N^A(\xi_N) \sigma(\vec y_\perp)\right]
\label{dipole}
\end{eqnarray}
in terms of the dipole cross section $\sigma(\vec y_\perp)$ and the
medium density $\rho_N^A(\xi_N)$. Using the short distance form or the leading
logarithmic approximation of the dipole cross
section \cite{Zakharov:1996fv},
\[
\rho_N^A(\xi_N)\sigma(\vec y_\perp)\approx \frac{1}{2}\hat q_F(\xi_N)y_\perp^2,
\]
one can obtain the expression in Eq.~(\ref{ydistr}) for the
transverse coordinate distribution $f(\vec y_\perp)$. One can
easily identify the first coefficient in the power expansion of
the dipole expansion $\hat q_F(\xi_N)$ with the quark transport
parameter as we have defined in our twist expansion approach
[Eq.~(\ref{fnlqhat})]. Therefore, the maximal two-gluon
correlation approximation for the dominant multiple gluon
correlation in nuclear medium in our study here is equivalent to
the dipole model approximation of the Wilson line approach
\cite{Wiedemann:2000za,Kovner:2003zj} when the short distance form
of the dipole cross section is used.

To relate our twist expansion result to the dipole model
approximation beyond the maximal two-gluon correlation, we 
can use the identity in Eq.~(\ref{lc-diff}) to recast the 
transverse coordinate distribution,
\begin{eqnarray}
f(\vec y_\perp)&=&\frac{1}{N_c}
\langle\!\langle{\rm Tr}\left[
e^{\vec y_\perp\cdot\partial_{\xi_\perp}}
{\cal L}^\dagger_\parallel(-\infty,\infty;\vec 0_\perp)
{\cal L}_\perp(-\infty;\vec 0_\perp,\vec \xi_\perp)
{\cal L}_\parallel(-\infty,\infty;\vec \xi_\perp)\right]
\rangle\!\rangle_{\xi_\perp=0}
=\frac{1}{N_c}\langle\!\langle {\rm Tr}
e^{-i\vec W_\perp(\infty)\cdot\vec y_\perp}\rangle\!\rangle,
\label{wilson}
\end{eqnarray}
in terms of the transport operator $\vec W_\perp(\infty)$
[Eq.~(\ref{transop})]. A Taylor expansion of the above
distribution in $\vec y_\perp$ in the extended adjoint two-gluon
correlation approximation will lead to a transverse
distribution as in Eq.~(\ref{hgtwst-dis}),
\begin{eqnarray}
f(\vec y_\perp)
\approx \exp\left\{-\frac{y_\perp^2}{4}\Delta_{F}(y_\perp^2)\right\}
=\exp\left\{-\frac{y_\perp^2}{4}\sum_{n=1}^{\infty}
\left(\frac{y_\perp^2}{4}\right)^{n-1}\frac{(-1)^{n-1}}{(n-1)!}
\Delta_{2nF}\right\}.
\label{diexpan}
\end{eqnarray}

Comparing the above distribution to the dipole model approximation 
in Eq.~(\ref{dipole}), one can relate the dipole cross section
to the nucleon TDD gluon distribution function,
\begin{equation}
\sigma(\vec y_\perp)\equiv y_\perp^2\frac{\pi^2 \alpha_s}{N_c}
[xf_g^N(x,y_\perp^2)]_{x=0}.
\end{equation}
This is exactly the cross section between a nucleon and a 
quark-anti-quark pair in a dipole configuration with transverse
separation $\vec y_\perp$ \cite{Huang:1997ii,Frankfurt:1993it}.

Our calculation can also be extended to a gluon
propagation.  The results are the same and one only has to
change the color factor to get the definition of the gluon
transport parameter,
\begin{equation}
\hat q_A(\xi_N,y_\perp^2)=\frac{4\pi^2\alpha_s C_A}{N_c^2-1}
\rho^A_N(\xi_N)[xf_g^N(x,y_\perp^2)]_{x= 0}.
\end{equation}
Comparison to the approximation of the averaged Wilson line,
\begin{equation}
\frac{1}{N_c^2-1}
\langle{\rm Tr}\left[
{\cal L}^\dagger_\parallel(-\infty,0;\vec 0_\perp)
{\cal L}_\parallel(-\infty,y^-;\vec y_\perp)\right]\rangle
\approx \exp\left[-\frac{1}{4}Q^2_{\rm sat}(x,y_\perp^2) y_\perp^2\right],
\end{equation}
in the study of gluon saturation in large nuclei \cite{Kovchegov:1998bi}
will relate the saturation scale $Q^2_{\rm sat}(y_\perp^2)$
with the path-integrated  gluon transport
parameter $\hat q_A(\xi_N,y_\perp^2)$,
\begin{equation}
Q^2_{\rm sat}(y_\perp^2)=\int d\xi_N^- \hat q_A(\xi_N,y_\perp^2)
=\frac{4\pi^2\alpha_s C_A}{N_c^2-1}
\int d\xi_N^- \rho^A_N(\xi_N) xf_g^N(x,y_\perp^2).
\end{equation}

The transverse scale dependence of the transport parameter, the
dipole cross section or the saturation scale in our calculation
come from contributions of higher-twist nucleon gluon
matrix elements and therefore are non-perturbative in nature. At
very short transverse distance scale, radiative corrections will
become important at the leading twist and they will give rise to 
a transverse scale dependence of the transport parameter that is 
governed by the Dokshitzer-Gribov-Lipatov-Altarelli-Parisi (DGLAP) 
evolution equations \cite{Dokshitzer:1977sg,Gribov:1972ri,Altarelli:1977zs}.

\subsection{Multiple gluon correlations in ${\cal N}=4$ SYM}

In the Taylor expansion of the TDD quark transport parameter
$\Delta_F(y_\perp^2)$ in Eq.~(\ref{diexpan}), the coefficients
are higher-twist nucleon gluon matrix elements which
generally involve multi-gluon correlations of the medium. 
Therefore studying these power corrections to the transport
parameter will shed light on multi-gluon correlations in a medium,
especially a strongly coupled system when they become important.

Non-perturbative calculation of the medium averaged Wilson line in
Eq.~(\ref{wilson}) is difficult for a strongly coupled system. The
developed technique of lattice QCD is not applicable because it is
formulated in the Euclidean space and is only suited for the study
of static thermodynamic observables. However, many transport
coefficients, such as the shear viscosity to entropy density ratio
$\eta/s$ \cite{Policastro:2001yc} and transverse momentum
broadening of a heavy quark
\cite{Gubser:2006nz,CasalderreySolana:2007qw}, have been studied
for ${\cal N}=4$ SYM theory in the large 't Hooft coupling
($\lambda\equiv g^2_{\rm SYM}N_c$) limit, employing the AdS/CFT
correspondence \cite{Maldacena:1997re}. Though SYM is not exactly
QCD, its study might provide indicative information on the
properties of a strongly coupled system.

Recently, the thermal averaged Wilson loop along the light-cone
with longitudinal distance $L^-$ and transverse
separation $y_\perp$ was calculated \cite{Liu:2006ug} in
the strong coupling limit of ${\cal N}=4$ SYM theory. The corresponding
transverse distribution from the calculated Wilson loop in the
fundamental representation is
\begin{equation}
f(\vec y_\perp)=\exp\left\{-a\sqrt{\lambda} L^- T
\left[\sqrt{\left(\frac{\pi T y_\perp}{2a}\right)^2+1}-1\right]\right\},
\label{ads1}
\end{equation}
where $T$ is the temperature and
$a=\sqrt{\pi}\Gamma(5/4)/\Gamma(3/4)\approx 1.311$. Note a factor
of $1/\sqrt{2}$ is missing here because of our definition of
light-cone variables. Expanding the exponent in a Taylor series of
the transverse distance $y_\perp^2$, one has
\begin{equation}
f(\vec y_\perp)=\exp\left[-\frac{y_\perp^2}{4} L^-
\frac{\sqrt{\lambda}\pi^2 T^3}{2a}
\sum_{n=1}^{\infty}\left(\frac{y_\perp^2}{4}\right)^{n-1}
\left(\frac{\pi^2 T^2}{2a^2}\right)^{n-1}
\frac{(-1)^{n-1}|(2n-3)!!|}{n!}\right].
\end{equation}
Comparing to Eq.~(\ref{diexpan}), one can extract the
leading-twist quark transport parameter in ${\cal N}=4$ SYM theory,
\begin{equation}
\hat q_F^{\rm SYM}=\frac{\sqrt{\lambda}\pi^2 T^3}{2a},
\end{equation}
which is half (versus 4/9 in QCD) of the gluon transport parameter
as obtained in Ref.~\cite{Liu:2006ug}.

Comparing the power corrections, one can also extract
higher-twist quark transport parameters in SYM(${\cal N}=4$),
\begin{equation}
\frac{\Delta_{2nF}^{\rm SYM}}{\Delta_{2F}^{\rm SYM}}=
\left(\frac{\pi^2T^2}{2a^2}\right)^{n-1}
\frac{|(2n-3)|!!}{n}\,\,  (n\ge 1).
\end{equation}
It is interesting to note that all higher-twist
gluon matrix elements that include multi-gluon
correlations are proportional to the leading-twist 
gluon matrix elements or two-gluon correlation. 
The coefficients are set only by the temperature of
the medium and are independent of the
coupling $\sqrt{\alpha_{\rm SYM}}$ and number of colors $N_c$.
This implies that multiple gluon correlations in the strong
coupling limit of SYM are as important as the two-gluon correlation.

Phenomenologically, one can use the relationship between the
transport parameter and the gluon distribution function in
Eq.~(\ref{tddq-gdis}) to obtain the
TDD gluon distribution density in a ${\cal N}=4$ SYM plasma,
\begin{equation}
\rho_N[xf_g(x,y_\perp^2)]_{x=0}^{\rm SYM}
=\frac{8aN_c^2}{\pi\sqrt{\lambda}} \frac{T}{y_\perp^2}
\left[\sqrt{\left(\frac{\pi T y_\perp}{2a}\right)^2+1}-1\right],
\end{equation}
which is proportional to $N_c^2$ for fixed t' Hooft couling
constant $\lambda$.

\section{Summary and Discussion}
\label{sec:summary}

In this paper, we have derived a gauge invariant form of nuclear
transverse momentum broadening distribution, utilizing the gauge
invariant TMD quark distribution function from
Ref.~\cite{Belitsky:2002sm}. We first express such TMD quark
distribution function in terms of a sum of all higher twist and
gauge invariant collinear parton matrix elements. These
higher-twist parton matrix elements are expectation values of the
moments of a transport operator $\vec W_\perp(y)$ which generates
the transverse momentum in a nucleon or nucleus when it acts on
the parton field $\psi(y)$. The defined transport operator $\vec
W_\perp(y)$ transforms like a covariant derivative and the final
expression is explicitly gauge invariant. With this general form
of the TMD parton distribution function, one can then calculate
any moments of the parton's transverse momentum in terms of the
higher twist parton matrix elements.

To calculate the nuclear broadening of transverse momentum
distribution, we approximate the nuclear matrix elements of $n$
pair of parton field operators as a product of $n$ nucleon parton
distributions, neglecting nuclear bounding effect and multiple
nucleon correlations which have non-leading nuclear size
dependence in a large nuclear. In other words, multiple gluon
correlations are assumed to be given as  products of two-gluon
correlations, which we called maximal two-gluon correlation
approximation. With such approximated nuclear matrix elements that
have the dominant nuclear size dependence of $A^{n/3}$ for fixed
dimension of the multi-parton operators, we can express the final
nuclear TMD quark distribution function in terms of the nucleon
TMD quark distribution. This form also obeys a 2-D diffusion
equation whose solution is a convolution of a Gaussian
distribution function and the nucleon TMD quark distribution. The
width of the Gaussian, or the mean total transverse momentum
broadening squared, is just the path integral of the quark
transport parameter $\hat q_F$, which is also defined in an
explicitly gauge invariant form and is related to the local gluon
distribution density.

Under an extended adjoint two-gluon correlation approximation,
one can resum some of the higher-twist nucleon gluon
matrix elements to obtain a transverse-distance-dependent
quark transverse parameter which is given by the TDD
gluon distribution function. Such a TDD quark transport
parameter will give rise to power corrections to the
Gaussian form of nuclear transverse momentum distribution
function.

We compared our final results with that of the Wilson line
approach to multiple parton
scattering \cite{Wiedemann:2000za,Kovner:2003zj}
in the dipole model approximation. The two results
are identical for short distance approximation of the
dipole cross section, which is equivalent to the maximal
two-gluon correlation approximation employed in our calculation.
If we relax the maximal two-gluon correlation to include
non-leading length-dependent contributions involving
higher-twist gluon distribution functions, one can then
relate the dipole cross section to the TDD gluon distribution
function in the medium.

We also compared our results with the AdS/CFT calculation \cite{Liu:2006ug}
of the transverse distribution of a Wilson line for a propagating quark in
the ${\cal N}=4$ SYM theory, in particular the power corrections
to the leading Gaussian distribution. We found that the SYM result
indicates the importance of multiple gluon correlations in a
strongly coupled system.

Though our final result for the nuclear modified transverse momentum
distribution contains all higher-twist collinear nuclear
parton matrix elements,
it is still only the leading twist contribution in terms of power suppression
${\cal O}(1/Q^n)$. One can follow the procedure as outlined in
Ref.~\cite{Liang:2006wp} to compute higher-twist contributions
to the momentum broadening which has the same nuclear length dependence
but are power suppressed in the momentum scale of the hard processes
as compared to the leading twist result obtained in this paper.

The transverse momentum broadening we calculated in this paper is
only valid in small transverse momentum region. At large transverse
momentum radiative corrections become important and will lead
to large logarithmic corrections to the transverse scale dependence
of the transport parameter \cite{jorge}. The form of nuclear broadening
will also have large power corrections to the Gaussian form.

The Gaussian form of the transverse momentum broadening in nuclear
medium we discussed in this paper will also have phenomenological
implications in the hard processes in hadron-nucleus and 
nucleus-nucleus collisions. Our study here justifies the phenomenological
approach to the initial transverse momentum 
broadening \cite{Wang:1998ww,Wang:2001cy}
in $p+A$ and $A+A$ collisions in which a Gaussian form of
the broadening is often used. Convoluted together with the
power-law-like transverse momentum spectra due to hard
scattering, the Gaussian broadening will naturally lead to
Cronin enhancement of the final hadron spectra at small and
intermediate transverse momentum.

\begin{center}
{\bf ACKNOWLEDGMENTS}
\end{center}

We would like to thank J.~Casalderrey-Solana for many useful discussions
about the dipole model and ${\cal N}=4$ SYM results. We also thank
A.~Majumder and F.~Yuan for helpful discussions during this work.
This work is supported  by the Director, Office of Energy
Research, Office of High Energy and Nuclear Physics, Division of
Nuclear Physics, of the U.S. Department of Energy under Contract No.
DE-AC02-05CH11231 and National Natural Science Foundation of China
under Project No. 10525523.

\section*{APPENDIX A: Transverse derivative of a gauge link}

\begin{figure}
\centerline{\includegraphics[width=14cm]{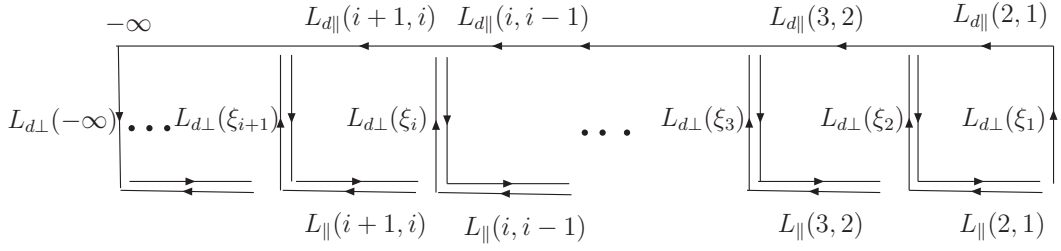}}
\caption{Splitting the cap-like gauge link into chains of closed
plaquette linked by short Wilson lines.} \label{fig2}
\end{figure}

Consider a cap-like gauge link,
\begin{equation}
{\cal L}_\sqcap\equiv
{\cal L}^{\dagger}_{d\perp}(-\infty)
{\cal L}_\parallel(-\infty,y^-;\vec y_\perp+d\vec y_\perp)
{\cal L}_{d\perp}(y^-)
\label{linkcap}
\end{equation}
with
\begin{equation}
{\cal L}_{d\perp}(y^-)\equiv
{\cal L}_\perp(y^-;\vec y_\perp+d\vec y_\perp,y_\perp,)
=1-igd\vec y_\perp \cdot \vec A_\perp(y^-,\vec y_\perp)
\end{equation}
and an infinitesimal transverse displacement $ d\vec y_\perp$. One
can break the longitudinal gauge link into a product of many
small ones each with an infinitesimal length $d\xi^-$,
\begin{eqnarray}
{\cal L}_\parallel(-\infty,y^-;\vec y_\perp+d\vec y_\perp)
&=&\cdots {\cal L}_{d\parallel}(i+1,i)\cdots
{\cal L}_{d\parallel}(3,2){\cal L}_{d\parallel}(2,1),
\label{linkd}
\\
{\cal L}_\parallel(-\infty,y^-;\vec y_\perp)
&=&\cdots {\cal L}_{\parallel}(i+1,i)\cdots
{\cal L}_{\parallel}(3,2){\cal L}_{\parallel}(2,1),
\end{eqnarray}
where ${\cal L}_{d\parallel}(i+1,i)$ and ${\cal L}_{\parallel}(i+1,i)$
are defined as
\begin{eqnarray}
{\cal L}_{d\parallel}(i+1,i)&\equiv&
{\cal L}_\parallel(\xi_{i+1}^-,\xi_i^-;\vec y_\perp+d\vec y_\perp),
\\
{\cal L}_\parallel(i+1,i)&\equiv&
{\cal L}_\parallel(\xi_{i+1}^-,\xi_i^-;\vec y_\perp),
\end{eqnarray}
and $\xi_i^-=y^- - (i-1)d\xi^-$.
Inserting unit matrices
\begin{equation}
1={\cal L}_{d\perp}(\xi_i){\cal L}_\parallel(i+1,i)
{\cal L}^\dagger_\parallel(i+1,i){\cal L}^\dagger_{d\perp}(\xi_i)
\end{equation}
between all neighboring links
${\cal L}_{d\parallel}(i+1,i)$ and ${\cal L}_{d\parallel}(i,i-1)$
in ${\cal L}_\parallel(-\infty,y^-,\vec y_\perp+d\vec y_\perp)$
[Eq.~(\ref{linkd})], as illustrated in Fig.~\ref{fig2},
except the last point where one instead inserts the unit matrix
\begin{equation}
1={\cal L}_\parallel(-\infty,-\infty+d\xi^-)
{\cal L}^\dagger_\parallel(-\infty,-\infty+d\xi^-)
\end{equation}
after ${\cal L}^\dagger_{d\perp}(-\infty)$.
One can then recast the cap-like gauge link,
\begin{eqnarray}
{\cal L}_\sqcap={\cal L}_\parallel(-\infty,-\infty+d\xi^-)
%{\cal L}_\Box(-\infty+d\xi^-)
\cdots {\cal L}_\parallel(i+1,i)
{\cal L}_\Box(\xi_i^-){\cal L}_\parallel(i,i-1)
{\cal L}_\Box(\xi_{i-1}^-)
\cdots
{\cal L}_\Box(\xi_2^-){\cal L}_\parallel(2,1)
{\cal L}_\Box(\xi_1^-),
\end{eqnarray}
as a product of closed plaquette,
\begin{equation}
{\cal L}_\Box(\xi_i^-)\equiv{\cal L}^\dagger_\parallel(i+1,i)
{\cal L}_{d\perp}^\dagger(\xi_{i+1}){\cal L}_{d\parallel}(i+1,i)
 {\cal L}_{d\perp}(\xi_{i}),
\end{equation}
that are linked by short Wilson lines ${\cal L}_\parallel(i+1,i)$.

Using the expansion of the closed plaquette \cite{khuang} up to
the linear term in $d\xi^- d\vec y_\perp$,
\begin{equation}
{\cal L}_\Box(\xi_i^-)=
1-igd\xi^- d\vec y_\perp\cdot \vec F_{+\perp}(\xi_i^-,\vec y_\perp)
\end{equation}
one can expand ${\cal L}_\sqcap$ up to the term linear in $d\vec y_\perp$,
\begin{eqnarray}
{\cal L}_\sqcap&=&{\cal L}_\parallel(-\infty,y^-;\vec y_\perp)
-igd\vec y_\perp\cdot \sum_{i=1}^{\infty} d\xi^-{\cal L}_\parallel(-\infty,\xi_i^-;\vec y_\perp)
\vec F_{+\perp}(\xi_i^-,\vec y_\perp)
{\cal L}_\parallel(\xi_i^-,y^-;\vec y_\perp)
\nonumber \\
&=&{\cal L}_\parallel(-\infty,y^-;\vec y_\perp)
-igd\vec y_\perp\cdot
\int_{-\infty}^{y^-} d\xi^-{\cal L}_\parallel(-\infty,\xi^-;\vec y_\perp)
\vec F_{+\perp}(\xi^-,\vec y_\perp)
{\cal L}_\parallel(\xi^-,y^-;\vec y_\perp).
\end{eqnarray}
Comparing to the direct expansion of ${\cal L}_\sqcap$
in Eq.~(\ref{linkcap}) in $d\vec y_\perp$,
\begin{eqnarray}
{\cal L}_\sqcap&=&{\cal L}_\parallel(-\infty,y^-;\vec y_\perp)
+d\vec y_\perp \cdot \left[
\vec\partial_{y_\perp}{\cal L}_\parallel(-\infty,y^-;\vec y_\perp)
\right.
\nonumber \\
&& -\left.
ig{\cal L}_\parallel(-\infty,y^-;\vec y_\perp) \vec A_\perp(y^-,\vec y_\perp)
+ig\vec A_\perp(-\infty,\vec y_\perp)
{\cal L}_\parallel(-\infty,y^-;\vec y_\perp) \right],
\end{eqnarray}
one obtains,
\begin{eqnarray}
\vec\partial_{y_\perp}{\cal L}_\parallel(-\infty,y^-;\vec y_\perp)&=&
-ig\vec A_\perp(-\infty,\vec y_\perp)
{\cal L}_\parallel(-\infty,y^-;\vec y_\perp)
+
{\cal L}_\parallel(-\infty,y^-;\vec y_\perp) \nonumber \\
&&\times\left[ig\vec A_\perp(y^-,\vec y_\perp)
-ig\int_{-\infty}^{y^-} d\xi^-{\cal L}^\dagger_\parallel(\xi^-,y^-;\vec y_\perp)
\vec F_{+\perp}(\xi^-,\vec y_\perp)
{\cal L}_\parallel(\xi^-,y^-;\vec y_\perp) \right].
\label{link-diff}
\end{eqnarray}

One can also obtain the above transverse derivative via the direct
expansion of the path-ordered longitudinal gauge link,
\begin{eqnarray}
&&\vec\partial_{y_\perp}{\cal L}_\parallel(-\infty,y^-;\vec y_\perp)
=\sum_{n=1}^{\infty}(-ig)^n\sum_{i=1}^n
\int^{-\infty}_{y^-}[d\xi]^n_1 A_+(\xi_1^-,\vec y_\perp)
 \cdots
\vec\partial_\perp A_+(\xi_i^-,\vec y_\perp)
\cdots A_+(\xi_n^-,\vec y_\perp)
\nonumber \\
&&=
\sum_{n=1}^{\infty}(-ig)^n\sum_{i=1}^n
\int^{-\infty}_{y^-}[d\xi]^n_1
 A_+(\xi_1^-,\vec y_\perp) \cdots
\left[\partial_+ \vec A_\perp(\xi_i^-,\vec y_\perp)
-\vec{\tilde F}_{+\perp}(\xi_i^-,\vec y_\perp)\right]
 \cdots A_+(\xi_n^-,\vec y_\perp) ,
\end{eqnarray}
where
\begin{equation}
\int^{-\infty}_{y^-}[d\xi]^n_1 \equiv
\int^{-\infty}_{y^-}d\xi_1\int_{y^-}^{\xi_1^-}d\xi_2^-
\cdots\int_{y^-}^{\xi_{n-1}^-}d\xi_n^-,
\end{equation}
\begin{eqnarray}
\vec{\tilde F}_{+\perp}(\xi_i^-,\vec y_\perp)
&\equiv&
\partial_+\vec A_\perp(\xi_i^-,\vec y_\perp)
-\vec\partial_\perp A_+(\xi_i^-,\vec y_\perp).
\end{eqnarray}
One can complete the integration of the terms with
$\partial_+\vec A_\perp(\xi_i^-,\vec y_\perp)$ by changing the
order of integration,
\begin{eqnarray}
\int_{y^-}^{\xi_{i-1}^-}d\xi_ i^-
\int_{y^-}^{\xi_i^-}d\xi_{i+1}^-
\partial_+ \vec A_\perp(\xi_i^-,\vec y_\perp)
&=&\int_{y^-}^{\xi_{i-1}^-}d\xi_{i+1}^-
\int_{\xi_{i+1}}^{\xi_{i-1}^-}d\xi_ i^-
\partial_+ \vec A_\perp(\xi_i^-,\vec y_\perp) \nonumber \\
&&\hspace{-2.in}=\int_{y^-}^{\xi_{i-1}^-}d\xi_{i+1}^-
\left[ \vec A_\perp(\xi_{i-1}^-,\vec y_\perp)
-\vec A_\perp(\xi_{i+1}^-,\vec y_\perp)\right],
\end{eqnarray}
for $i=1,2, \dots,n-1$, with $\xi_0^-=-\infty$ and
\begin{equation}
\int_{y^-}^{\xi_{n-1}^-}d\xi_ n^-
\partial_+ \vec A_\perp(\xi_n^-,\vec y_\perp)
=\vec A_\perp(\xi_{n-1}^-,\vec y_\perp)
-\vec A_\perp(y^-,\vec y_\perp).
\end{equation}
We can rearrange the sum of terms associated with
$\partial_+\vec A_\perp(\xi_i^-,\vec y_\perp)$ in the
$n$th order (in the coupling $g$) of the expansion as,
\begin{eqnarray}
&&\sum_{i=1}^n
\int^{-\infty}_{y^-}[d\xi]^n_1
 A_+(\xi_1^-,\vec y_\perp) \cdots
\partial_+ \vec A_\perp(\xi_i^-,\vec y_\perp)
\cdots A_+(\xi_n^-,\vec y_\perp)
, \nonumber\\
%%%%%%%%%%%
&&=\int^{-\infty}_{y^-}[d\xi]^{n-1}_1
\left[\prod_{j=1}^{n-2} A_+(\xi_j^-,\vec y_\perp)\right]
A_+(\xi_{n-1}^-,\vec y_\perp)
\left[\vec A_\perp(\xi_{n-1}^-,\vec y_\perp)
-\vec A_\perp(y^-,\vec y_\perp)\right]
\nonumber \\
&&\hspace{0.5in}+\sum_{i=2}^{n-1}
\int^{-\infty}_{y^-}[d\xi]^{i-1}_1
\left[\prod_{j=1}^{i-2}A_+(\xi_j^-,\vec y_\perp)\right]
\int^{\xi_{i-1}^-}_{y^-}d\xi_{i+1}^-
\nonumber \\
&&\times
A_+(\xi_{i-1}^-,\vec y_\perp)
\left[ \vec A_\perp(\xi_{i-1}^-,\vec y_\perp)
-\vec A_\perp(\xi_{i+1}^-,\vec y_\perp)\right]
A_+(\xi_{i+1}^-,\vec y_\perp)
\int^{\xi_{i+1}}_{y^-}[d\xi]^{n}_{i+2}
\prod_{j=i+2}^nA_+(\xi_j^-,\vec y_\perp)
\nonumber \\
%%%%
&&\hspace{0.5in}+ \int^{-\infty}_{y^-}[d\xi]^{n}_2
\left[\vec A_\perp(-\infty,\vec y_\perp)
-\vec A_\perp(\xi_2^-,\vec y_\perp)\right]
A_+(\xi_2^-,\vec y_\perp)
\prod_{j=3}^{n} A_+(\xi_j^-,\vec y_\perp)
\nonumber \\
%%%%
&&=-\int^{-\infty}_{y^-}[d\xi]^{n-1}_1
\left[\prod_{j=1}^{n-1}A_+(\xi_j^-,\vec y_\perp)\right]
\vec A_\perp(y^-,\vec y_\perp)
+\vec A_\perp(-\infty,\vec y_\perp)
\int^{-\infty}_{y^-}[d\xi]^{n}_2
\prod_{j=2}^n A_+(\xi_j^-,\vec y_\perp)
%%%%
\nonumber \\
&&+\sum_{i=1}^{n-1}
\int^{-\infty}_{y^-}[d\xi]^{i-1}_1
\left[\prod_{j=1}^{i-1}A_+(\xi_j^-,\vec y_\perp)\right]
\int^{\xi_{i-1}^-}_{y^-}d\xi_{i}^-
\left[A_+(\xi_i^-,\vec y_\perp),\vec A_\perp(\xi_i^-,\vec y_\perp)\right]
\int^{\xi_{i}}_{y^-}[d\xi]^{n}_{i+1}
\prod_{j=i+1}^{n-1}A_+(\xi_j^-,\vec y_\perp).
\end{eqnarray}
The terms containing the commutator,
$[A_+(\xi_i^-,\vec y_\perp),\vec A_\perp(\xi_i^-,\vec y_\perp)]$,
can be combined with $\vec{\tilde F}_{+\perp}(\xi_i^-,\vec y_\perp)$
in the $(n-1)$th order of the expansion to give the gluon field strength
tensor,
\begin{eqnarray}
\vec F_{+\perp}(\xi_i^-,\vec y_\perp)
&=&\vec{\tilde F}_{+\perp}(\xi_i^-,\vec y_\perp)
+ig [A_+(\xi_i^-,\vec y_\perp),\vec A_\perp(\xi_i^-,\vec y_\perp)].
\end{eqnarray}
Note that
\begin{eqnarray}
\sum_{n=1}^{\infty}(-ig)^n
\int^{\xi_{i}}_{y^-}[d\xi]^{n}_{i+1}
\prod_{j=i+1}^{n-1}A_+(\xi_j^-,\vec y_\perp)
={\cal L}_\parallel(\xi_i,y^-,\vec y_\perp),
\end{eqnarray}
and
\begin{eqnarray}
&&\sum_{i=1}^{\infty}(-ig)^{i-1}
\int^{-\infty}_{y^-}[d\xi]^{i-1}_1
\left[\prod_{j=1}^{i-1}A_+(\xi_j^-,\vec y_\perp)\right]
\int^{\xi_{i-1}^-}_{y^-}d\xi_{i}^-
\nonumber \\
&&=\sum_{i=1}^{\infty}(ig)^{i-1}
\int^{-\infty}_{y^-}d\xi_{i}^-
\int_{-\infty}^{\xi_i^-}d\xi_{i-1}^-
\cdots
\int_{-\infty}^{\xi_2^-}d\xi_1^-
A_+(\xi_1^-,\vec y_\perp)\cdots A_+(\xi_{i-1}^-,\vec y_\perp)
\nonumber \\
&&=\int^{-\infty}_{y^-}d\xi_{i}^-
{\cal L}^\dagger_\parallel(\xi_i,-\infty,\vec y_\perp).
\end{eqnarray}
One obtains now the transverse derivative of the longitudinal gauge link,
\begin{eqnarray}
\vec\partial_{y_\perp}{\cal L}_\parallel(-\infty,y^-;\vec y_\perp)&=&
-ig\int_{-\infty}^{y^-}d\xi
{\cal L}^\dagger_\parallel(\xi^-,-\infty)
\vec{F}_{+\perp}(\xi^-,\vec y_\perp)
{\cal L}_\parallel(\xi^-,y^-;\vec y_\perp)
\nonumber \\
&&\hspace{0.5in}
+ig {\cal L}_\parallel(-\infty,y^-;\vec y_\perp)\vec A_\perp(y^-,\vec y_\perp)
-ig\vec A_\perp(-\infty,\vec y_\perp)
{\cal L}_\parallel(-\infty,y^-;\vec y_\perp).
\end{eqnarray}

Another general form of the above identity is
\begin{eqnarray}
\vec D_\perp(y_1^-,\vec y_\perp){\cal L}_\parallel(y_1^-,y^-;\vec y_\perp)&=&
-ig\int_{y_1^-}^{y^-}d\xi
{\cal L}^\dagger_\parallel(\xi^-,y_1^-)
\vec{F}_{+\perp}(\xi^-,\vec y_\perp)
{\cal L}_\parallel(\xi^-,y^-;\vec y_\perp)
\nonumber \\
&&\hspace{0.5in}
+{\cal L}_\parallel(y^-_1,y^-;\vec y_\perp)
\vec D_\perp(y^-,\vec y_\perp).
\label{coidnty}
\end{eqnarray}

Using the above identity,
the derivative operation on the gauge link in the TMD quark
distribution function,
\begin{equation}
{\cal L}_{\rm TMD}(0,y)\equiv
{\cal L}^\dagger_\parallel (-\infty,0;\vec 0_\perp)
{\cal L}^\dagger_\perp(-\infty;\vec y_\perp,\vec 0_\perp)
{\cal L}_\parallel (-\infty,y^-;\vec y_\perp),
\end{equation}
will yield,
\begin{eqnarray}
\vec\partial_{y_\perp}{\cal L}_{\rm TMD}(0,y)
&=&{\cal L}_{\rm TMD}(0,y)\vec\partial_{y_\perp}
+{\cal L}^\dagger_\parallel (-\infty,0;\vec 0_\perp)
{\cal L}^\dagger_\perp(-\infty;\vec y_\perp,\vec 0_\perp)
\vec\partial_{y_\perp}{\cal L}_\parallel(-\infty,y^-;\vec y_\perp)
\nonumber \\
&&\hspace{0.5in}+ig {\cal L}^\dagger_\parallel (-\infty,0;\vec 0_\perp)
{\cal L}^\dagger_\perp(-\infty;\vec y_\perp,\vec 0_\perp)
\vec A_\perp(-\infty,\vec y_\perp)
{\cal L}_\parallel(-\infty,y^-;\vec y_\perp) \nonumber \\
&&\hspace{-1.0in}=
{\cal L}_{\rm TMD}(0,y)
\left[\vec D_\perp(y^-,\vec y_\perp)
-ig\int_{-\infty}^{y^-} d\xi^-{\cal L}^\dagger_\parallel(\xi^-,y^-;\vec y_\perp)
\vec F_{+\perp}(\xi^-,\vec y_\perp)
{\cal L}_\parallel(\xi^-,y^-;\vec y_\perp) \right].
\label{lc-diff}
\end{eqnarray}

If we define
\begin{equation}
\vec {\cal F}_\perp(y^-,\vec y_\perp)
\equiv g\int_{-\infty}^{y^-}
d\xi^-{\cal L}^\dagger_\parallel(\xi^-,y^-;\vec y_\perp)
\vec F_{+\perp}(\xi^-,\vec y_\perp)
{\cal L}_\parallel(\xi^-,y^-;\vec y_\perp),
\end{equation}
it is easy to use Eq.~(\ref{coidnty}) to show
\begin{equation}
D_{\perp i}(y,\vec y_\perp) \vec {\cal F}_\perp(y^-,\vec y_\perp)
=\vec {\cal F}_\perp(y^-,\vec y_\perp) D_{\perp i}(y,\vec y_\perp)
+g\int_{-\infty}^{y^-} d\xi^-{\cal L}^\dagger_\parallel(\xi^-,y^-;\vec y_\perp)
D_{\perp i}^{A}\vec F_{+\perp}(\xi^-,\vec y_\perp)
{\cal L}_\parallel(\xi^-,y^-;\vec y_\perp),
\label{adjco}
\end{equation}
where $\vec D_\perp^AF=\vec\partial_\perp+ig[\vec A_\perp,F]$ is the
covariant derivative in the adjoint representation.

\section*{APPDENIX B: Extended two-gluon correlation approximation}
\label{sec:appb}

In this Appendix we will examinehigher-twist nucleon gluon matrix elements
that are neglected in the extended two-gluon correlation approximation
in Sec.~\ref{sec:tmdgluon}.

In the connected part of the twist-four gluon matrix elements
one can extend the product of transport operator without assuming
extended two-gluon correlation approximation,
\begin{eqnarray}
\Delta_{4F}&=&\frac{1}{2N_c} \langle\!\langle{\rm Tr}
\left[W_\perp(y^-)\right]^4
\rangle\!\rangle_{AC}
\nonumber \\
&=&\frac{1}{2N_c}\langle\!\langle{\rm Tr}
\left[{\cal F}_\perp\right]^4
\rangle\!\rangle_{AC}
+\frac{2}{2N_c}\langle\!\langle{\rm Tr}
\left[(i\vec D_\perp^A\cdot\vec{\cal F}_\perp){\cal F}_\perp^2
+\vec{\cal F}_\perp\cdot(i\vec D_\perp^A\cdot\vec{\cal F}_\perp)
\vec{\cal F}_\perp+{\cal F}_\perp^2(i\vec D_\perp^A\cdot\vec{\cal F}_\perp)
\right]\rangle\!\rangle_{A}
\nonumber \\
&-&\frac{1}{2N_c}\langle\!\langle{\rm Tr}
\left[3\vec{\cal F}_\perp\cdot({D_\perp^A}^2 \vec{\cal F}_\perp)
+({D_\perp^A}^2\vec{\cal F}_\perp)\cdot \vec{\cal F}_\perp
+3(D_\perp^A\cdot\vec{\cal F})(D_\perp^A\cdot\vec{\cal F}_\perp)
\right]\rangle\!\rangle_{A}
\nonumber \\
&=&\frac{1}{4N_c} \langle\!\langle
\vec {\cal F}_\perp^a\left[\left({\cal F}_\perp^O\right)^2_{ab}
+(iD_\perp^A)^2_{ab}
+6(i\vec D_\perp^A\cdot\vec{\cal F}_\perp^O)_{ab}\right]
\vec {\cal F}_\perp^b \rangle\!\rangle_A.
\end{eqnarray}
Terms linear in ${\cal F}_\perp(y^-)$ are dropped since they
vanish after the medium average. The following identities along 
medium averaged matrix elements are also used,
\begin{eqnarray}
\langle\!\langle {\rm Tr}\left[ \vec{\cal F}_{\perp}\cdot(D_\perp^A)^2
\vec{\cal F}_{\perp} \right]\rangle\!\rangle_A
=\langle\!\langle {\rm Tr}\left[ (D_\perp^A)^2\vec{\cal F}_{\perp}
\cdot\vec{\cal F}_{\perp} \right]\rangle\!\rangle_A
=-\langle\!\langle {\rm Tr}\left[ (\vec D_\perp^A \cdot
\vec{\cal F}_{\perp})\cdot
(\vec D_\perp^A \cdot\vec{\cal F}_{\perp})\right] \rangle\!\rangle_A,
\end{eqnarray}
\begin{eqnarray}
\langle\!\langle {\rm Tr}\left[ {\cal F}_{\perp}^2 i\vec D_\perp^A
\cdot {\cal F}_{\perp} \right]\rangle\!\rangle_A
=\langle\!\langle {\rm Tr}\left[ \vec{\cal F}_{\perp}
\cdot (i\vec D_\perp^A \cdot {\cal F}_{\perp})
\vec{\cal F}_{\perp}\right] \rangle\!\rangle_A
=\langle\!\langle {\rm Tr}\left[ i\vec D_\perp^A\cdot \vec {\cal F}_{\perp}
{\cal F}_{\perp}^2 \right]\rangle\!\rangle_A.
\end{eqnarray}
One can see that there are differences between the higher-twist
gluon matrix elements in the above expansion and that under
extended two-gluon correlation approximation 
in Sec.\ref{sec:tmdgluon} which include some extra higher-twist
gluon matrix elements.

\end {document}